\documentclass[12pt]{article}
\usepackage{amsthm, amsmath}
\usepackage[titletoc,toc,title]{appendix}

\usepackage{graphicx}
\usepackage[round]{natbib}
\usepackage{enumerate}
\usepackage{threeparttable}
\usepackage{lscape}
\usepackage{multirow}
\usepackage[singlelinecheck=off]{caption}
\usepackage{array}
\usepackage[a4paper, total={6in, 9in}]{geometry}

\DeclareMathOperator*{\argmax}{arg\,max}
\newtheorem{thm}{Theorem}

\newcommand{\E}{\mathrm{E}}

\newcommand{\BIC}{\text{BIC}}
\newcommand{\tr}{\text{tr}}
\newcommand{\tbeta}{\boldsymbol{\tilde \beta}}
\newcommand{\hbeta}{\boldsymbol{\hat \beta}}
\newcommand{\bbeta}{\boldsymbol{\beta}}
\newcommand{\boldeta}{\boldsymbol{\eta}}
\newcommand{\bZ}{\mathbf{Z}}
\newcommand{\bT}{\mathbf{T}}
\newcommand{\bC}{\mathbf{C}}
\newcommand{\bz}{\mathbf{z}}
\newcommand{\bu}{\mathbf{u}}

\newcommand{\bzero}{\mathbf{0}}
\newcommand{\bP}{\mathbf{P}}
\newcommand{\bb}{\mathbf{b}}

\newcommand{\bU}{\mathbf{U}}
\newcommand{\bSigma}{\boldsymbol{\Sigma}}
\newcommand{\bOmega}{\boldsymbol{\Omega}}
\newcommand{\B}{\mathcal{B}}
\newcommand{\szero}{s^{(0)}}
\newcommand{\sone}{\mathbf{s}^{(1)}}
\newcommand{\stwo}{\mathbf{s}^{(2)}}
\newcommand{\spp}{\mathbf{s}^{(p)}}
\newcommand{\Szero}{S^{(0)}}

\newcommand{\Sp}{\mathbf{S}^{(p)}}

\newcommand{\vt}{\mathbf{v}}
\newcommand{\hw}{\hat{w}}
\newcommand{\hG}{\hat{G}}
\newcommand{\hS}{\hat{S}}
\newcommand{\bD}{\mathbf{D}}
\newcommand{\bA}{\mathbf{A}}
\newcommand{\bv}{\mathbf{v}}
\newcommand{\by}{\mathbf{y}}
\newcommand{\hSzero}{\hat{S}^{(0)}}
\newcommand{\hSone}{\hat{\mathbf{S}}^{(1)}}
\newcommand{\hStwo}{\hat{\mathbf{S}}^{(2)}}
\newcommand{\hSp}{\hat{\mathbf{S}}^{(p)}}
\newcommand{\bI}{\mathbf{I}}

\newcolumntype{L}{>{$}l<{$}}
\newcolumntype{C}{>{$}c<{$}}
\newcolumntype{R}{>{$}r<{$}}

\newcommand{\footremember}[2]{%
    \footnote{#2}
    \newcounter{#1}
    \setcounter{#1}{\value{footnote}}%
}
\newcommand{\footrecall}[1]{%
    \footnotemark[\value{#1}]%
} 

\begin{document}

\title{Penalized Variable Selection for Multi-center Competing Risks Data}

\author{Zhixuan Fu   \and
        Shuangge Ma \and
        Haiqun Lin \and
        Chirag R Parikh\and
        Bingqing Zhou
          %etc.
}

\author{%
  Zhixuan Fu\footremember{yale}{Biostatistics Department, Yale University, 60 College Street, New Haven, CT 06510, U.S.}
  \and Shuangge Ma \footrecall{yale}%
  \and Haiqun Lin\footrecall{yale} %
  \and Chirag R Parikh \footremember{cp}{Section of Nephrology, Department of Internal Medicine, Yale University, 60 Temple Street, Suite 6C, New Haven, CT 06510, U.S.}
  \and Bingqing Zhou \footremember{bz}{ bingqing.zhou@yale.edu}
  }
\date{}

\maketitle

\begin{abstract}
We consider variable selection in competing risks regression for multi-center data. 
Our research is motivated by deceased donor kidney transplants, from which recipients would experience graft failure, death with  functioning graft (DWFG), or graft survival. The occurrence of DWFG precludes graft failure from happening and therefore is a competing risk. Data within a transplant center may be correlated due to a latent center effect, such as varying patient populations, surgical techniques, and patient management. 
The proportional subdistribution hazard (PSH) model has been frequently used in the regression analysis of competing risks data. Two of its extensions, the stratified and the marginal PSH models, can be applied to multi-center data to account for the center effect. In this paper, we propose penalization strategies for the two models, primarily to select important variables and estimate their effects whereas correlations within centers serve as a nuisance. Simulations demonstrate good performance and computational efficiency for the proposed methods. It is further assessed using an analysis of data from the United Network of Organ Sharing.

\end{abstract}
\section{Introduction}
\label{sec1}
Kidney transplant is the most cost-effective therapy for patients with end-stage renal disease, prolonging survival and improving quality of life. This practice is undergoing a tremendous shortage of organs. As of September, 2015, over 100,000 patients are on the waiting list in the United States \citep{unos}. From 2004 to 2014, the annual number of kidney transplants barely changed, ranging from 16,000 to 17,000 \citep{unos}, which is greatly disproportionate to the demand. Expanded donor criteria kidneys from older and sicker donors are being procured as a reaction to the shortage, which has led to greater discard rates of donated kidneys as well as more complications for recipients, including shorter graft survival  \citep{ECD}. Proper allocation of kidneys is more important than ever.

The Organ Procurement and Transplantation Network (OPTN) approved a new national deceased donor kidney allocation policy in 2013 in reaction to this shortage, which prioritizes candidates with longer estimated life expectancy to receive kidneys with the potential to function longer.
The Kidney Donor Risk Index (KDRI)  \citep{rao} was adopted to measure graft survival of deceased donor kidneys. Graft survival is defined as the time from the day of transplantation to the earliest onset of graft failure or death. A lower KDRI is associated with longer graft survival. This index includes 10 donor factors, which are identified by a variable selection procedure that sequentially eliminates insignificant factors. 

Two complications are present when the primary interest is to specifically predict the risk of graft failure, as opposed to the composite risk of graft failure and death as in the KDRI, using multi-center data. One is the presence of competing risks.
After kidney transplantations, recipients may experience graft survival, graft failure, or death with functioning graft (DWFG). Those who have died with a functioning graft can no longer develop graft failure. DWFG is therefore a competing risk to graft failure. 
The other complication is the center effect arising from available data of deceased donor kidney transplants. The data were collected by an ongoing organ transplant registry of the United Network of Organ Sharing (UNOS) from all transplant centers in the US. Patients within a center may have correlated outcomes due to a latent center effect, such as surgical techniques, patient management, and patient characteristics \citep{center1, center2, center3}. 
We therefore propose penalized variable selection strategies that account for the presence of competing risks and the center effect. Selected factors can be used to construct prognostic models for predicting the risk of graft failure.

In competing risks regression, the proportional subdistribution hazard (PSH) model  \citep{finegray} has become popular for its direct assessment of covariate effects on the cumulative incidence function. 
Some of its extensions can  account for within-center correlations.  \cite{pshfrailty} and \cite{HLfrailty} proposed a PSH frailty model treating the center effect as a random sample from a distribution. 
\cite{hafrailty} extended this model to handle potential heterogeneity in the treatment effect among centers.
\cite{zhoustratified} developed a stratified approach to account for varying patient populations by assuming an unspecified center-specific baseline subdistribution hazard. \cite{zhoucluster} proposed a marginal PSH model to estimate effects of covariates on the marginal cumulative incidence function under a working independence assumption; the variance estimator was adjusted to accommodate the within-center correlations.

Several variable selection methods have been extended to the competing risks setting.  \cite{kuk} extended stepwise selection to the PSH model, by developing selection criteria based on Akaike Information Criteria (AIC), Bayesian Information Criteria (BIC) and a modified BIC for competing risks data (BICcr). Although simple and easy to use,  stepwise selection is computationally intensive and unstable, and its theoretical properties are largely unknown \citep{scad, fu}.  \cite{fu} developed a generalized penalized variable selection methodology for the PSH model and established the asymptotic properties for the penalized estimator. However, the stepwise selection and the penalized PSH model cannot account for the center effect.
%Theses methods can be applied to the fixed effects PSH model for selecting covariates while keeping center indicators in the model. When the number of centers is large relative to center size, numerical problems may occur, owning to the large number of parameters for center effects. When all event times in one center are smaller or larger that event times in another center, then the estimated regression coefficient for center indicator does not converge. Instead, the estimate is $-\infty$ or $+\infty$, respectively.
\cite{haselection} proposed a variable selection procedure that penalizes a hierarchical likelihood of the PSH frailty model. Their method assumes frailties follow a log-normal distribution and simultaneously perform variable selection and estimation of covariate effects and frailties. Efficiency in standard errors is gained if the distribution of frailties is approximately true  \citep{frailtyerror}, but potentially misleading if misspecified. If the study interest only lies in variable selection and estimation of covariate effects, introducing parameters to explicitly model center effects either through indicator variables or frailties does not seem to be worthwhile, because it complicates the analysis of covariates effects and may be subject to numerical difficulties or bias under violations of model assumptions.

In this paper, we propose to extend the penalized variable selection method  \citep{fu} to the stratified and the marginal PSH models, treating within-center correlations as a nuisance. The method for the stratified PSH model can be applied to data exhibiting two types of stratification regimes: the regularly stratified and the highly stratified. The former has a small number of large centers and the latter has a large number of small centers. Center effects are modeled through center-specific subdistribution hazards.
The method for the marginal PSH model is suitable for data with a large number of centers. The proposed variance estimator accommodates the correlation within centers.
We also briefly describe the extension of the proposed methods to group variable selection, which can select pre-specified groups of variables collectively.

The remainder of this paper is organized as follows. In Section \ref{sec2}, we present the proposed penalization methods and address implementation issues.  In Section \ref{sec3}, simulations are conducted to evaluate the performance of the proposed methods. Section \ref{sec4} applies the methods to the data from the UNOS. Section \ref{sec5} contains discussion.

\section{Penalized Variable Selection for Multi-center Competing Risks Data}
\label{sec2}
\subsection{Notation}
\label{sec2.1}
Suppose there are $K$ centers with $n_k$ patients for center k, $k=1, \dots, K$. The total number of observations is $\Sigma_{k=1}^Kn_k=n$.  Cause 1 of failure is graft failure and cause 2 is the competing risk, DWFG.
For patient $i$ within center $k$, denote the failure time, the censoring time, and the failure cause as $T_{ki}$, $C_{ki}$,  and $\epsilon_{ki} \in \{1,2\}$. Let $\bZ_{ki} = \{Z_{1ki}, \dots, Z_{dki}\}$ be a $d \times 1$ vector of covariates. We observe $\{X_{ki}, \delta_{ki}, \delta_{ki}\epsilon_{ki}, \bZ_{ki}, \xi_{ki}\}$, where $X_{ki}=$min$(T_{ki}, C_{ki})$,  $\delta_{ki}=I(T_{ki}\leq C_{ki})$ is the event indicator, and $\xi_{ki} \in \{1, \dots, K\}$ is a center indicator. 

Let $\bT_k$ be a vector of $(T_{k1}, \dots, T_{kn_k})^T$, and $\boldsymbol{\epsilon}_k$, $\bC_k$, and $\bZ_k$ are defined similarly. We assume that $(\bT_k, \boldsymbol{\epsilon}_k, \bC_k, \bZ_k)$ are independent and identically distributed; $(\bT_k, \boldsymbol{\epsilon}_k)$ and $\bC_k$ are independent given $\bZ_k$. Patients within center $k$ are assumed to be exposed to a common center effect, which implies components of $(\bT_k, \boldsymbol{\epsilon}_k)$ may be correlated conditional on $\bZ_k$.

%Denote $v_k$ as the frailty for center $k$, which is a random variable summarizing the factors shared among patients within center $k$. Due to this shared frailty, failure times for patients within a center are correlated.

%%%%%%%%%%%%%%%%%%%%%%%%%%%%%%%%%%%%%%%%%%%
%%%%%%%%%%%%%%%%%%%%%%%%%%%%%%%%%%%%%%%%%%%
\subsection{Penalized Stratified PSH Model}
\label{sec2.2}
\subsubsection{Model}
\label{model s}
The stratified PSH model \citep{zhoustratified} is a conditional (or center-specific) approach that models the subdistribution hazards for graft failure in each center separately. It can be regarded as an analogy of the stratified Cox model for the competing risks setting. For center $k$, the cumulative incidence function for graft failure is defined as $F_{1k}(t;\bZ_{ki})=Pr(T_{ki}\leq t, \epsilon_{ki}=1|\bZ_{ki}, \xi_{ki}=k)$. The corresponding subdistribution hazard is $\lambda_{1k}(t;\bZ_{ki})=dF_{1k}(t; \bZ_{ki})/\{1-F_{1k}(t; \bZ_{ki})\}$.  
Assuming proportional subdistributional hazards, $\lambda_{1k}(t; \bZ_{ki})$ can be written as
\begin{equation}
\label{strat subhaz}
\lambda_{1k}(t; \bZ_{ki})=\lambda_{1k0}(t)\exp(\boldsymbol{\beta}^T\bZ_{ki}),
\end{equation}
where $\lambda_{1k0}(t)$ is an unspecified center-specific baseline subdistribution hazard, and $\boldsymbol{\beta}$ is a $d\times 1$ vector of regression coefficients. Within-center correlation arises from all patients being exposed to a common center effect.

The log-partial likelihood function is a linear combination of the likelihood for every center and is defined as 
\begin{equation}
\label{strat likelihood}
\begin{aligned}
l^S(\bbeta) &=\sum_{k=1}^K l_k(\bbeta) \\
& =\sum_{k=1}^K\sum_{i=1}^{n_k}\delta_{ki}I(\epsilon_{ki}=1)\{\beta^T\bZ_{ki}-\log \sum_{i'=1}^{n_k}\hw_{ki'}(X_{ki})Y_{ki'}(X_{ki})\exp(\beta^T\bZ_{ki'})\},
\end{aligned}
\end{equation}
where $l_k(\bbeta)$ is the log-partial likelihood for center $k$, $Y_{ki}(t)=1-I(T_{ki}\leq t-, \epsilon_{ki}=1)$ indicates whether the patient is still at risk of graft failure, and $\hw_{ki}(t)$ is a time-dependent weight for patient $i$ within center $k$.  Under the two stratification regimes, $\hw_{ki}(t)$ is defined differently using the inverse probability of censoring
weighting techniques \citep{robins1992recovery, zhoustratified} to achieve unbiased estimation. For regularly stratified data, observations are assumed to be i.i.d. within centers and the censoring distribution is center-dependent. In this case, the weight is defined as $\hw_{ki}(t)=I(C_{ki}\geq T_{ki}\land t)\hG_k(t)/\hG_k(X_{ki}\land t)$, where $G_k(t)=Pr(C_{ki}\geq t), i=1, \dots, n_k$, and $\hat G_k(.)$ is the Kaplan-Meier estimator of $G_k(.)$. It is possible to estimate $G_k(.)$ from a Cox model by including some potentially important covariates on the censoring. For highly stratified data, observations and censoring are assumed to be i.i.d. across centers, leading to a weight of $\hw_{ki}(t)=I(C_{ki}\geq T_{ki}\land t)\hG(t)/\hG(X_{ki}\land t)$, where $G(t)=Pr(C_{ki}\geq t)$ is the common censoring distribution, $i=1, \dots, n_k, k=1, \dots, K$.
By time $t$, if a patient has not experienced any event, $\hw_{ki}(t)Y_{ki}(t)=1$; if the patient is right censored or has experienced graft failure, $\hw_{ki}(t)Y_{ki}(t)=0$; if the patient has experienced DWFG, $\hw_{ki}(t)Y_{ki}(t)=\hG_k(t)/\hG_k(X_{ki})$ or $\hG(t)/\hG(X_{ki})$ ranging from 0 to 1. Covariate effects are estimated through maximizing function (\ref{strat likelihood}) and relative risks obtained from the covariates compare two patients within the same center \citep{david}.  

To perform variable selection and parameter estimation simultaneously, we use a penalized partial likelihood for the model in equation (\ref{strat subhaz}), which is defined as 
\begin{equation}
Q^S(\bbeta)=\sum_{k=1}^K \{l_k(\bbeta)-n_k\sum_{j=1}^dp_{\lambda}(|\beta_j|)\}=l^S(\bbeta)-n \sum_{j=1}^dp_{\lambda}(|\beta_j|),
\label{objective func s}
\end{equation}
where $p_{\lambda}(|\beta_j|)$ is a penalty function with tuning parameter $\lambda$, controlling the complexity of selected models. A larger $\lambda$ tends to choose a simpler model with fewer selected variables. Function (\ref{objective func s}) is essentially penalizing the log-partial likelihood of the ordinary PSH model for every center, and the center-specific penalty is $n_k\sum_{j=1}^dp_{\lambda}(|\beta_j|)$. 
The penalized estimator, denoted as $\tbeta^S$, is the maximizer of the objective function (\ref{objective func s}),
\begin{equation*}
\label{strat obj max}
\tbeta^S=\argmax Q^S(\bbeta).
\end{equation*}
Since the proportional hazard specification is conditional on centers, $\lambda_{1k0}(t)$ has ``summed out" in  $l_k(\bbeta)$. The objective function $Q^S(\bbeta)$ does not contain the center-specific baseline subdistribution hazards, which incorporate the center effect. Thus, the estimation of $\tbeta^S$ is not changed by the center effect.

Various penalty functions have been proposed in other settings; see  \cite{overview} for an extensive review. We consider four popular  penalties in this paper, but our results can be extended to other penalties: 
\begin{enumerate}[(a).]
\item Least Absolute Shrinkage and Selection Operator (LASSO) \citep{lasso}:  $p_{\lambda}(|\beta_j|)=\lambda|\beta_j|$. 
\item Adaptive LASSO (ALASSO) \citep{alasso}: $p_{\lambda}(|\beta_j|)=\lambda \theta_j|\beta_j|$, where $\theta_j$ is a data-adaptive weight assigned to $\beta_j$. We use $\theta_j=|\hat\bbeta^S_j|^{-1}$, where $\hat\bbeta^S$ is the maximizer of the log-partial likelihood function. 
\item Smoothly Clipped Absolute Deviation (SCAD) \citep{scad}: $p'_{\lambda}(|\beta_j|, \alpha)=\lambda I(|\beta_j|\leq \lambda)+\frac{(\alpha\lambda-|\beta_j|)_{+}}{(\alpha-1)}I(|\beta_j|>\lambda)$ and $\alpha>2$ is a tuning parameter and $(x)_{+}$ indicates the positive part of $x$. 
\item Minimax Concave Penalty (MCP) \citep{mcp}: $p'_{\lambda}(|\beta_j|, \gamma)=(\lambda-\frac{|\beta_j|}{\gamma})_+$ and $\gamma>1$ is a tuning parameter.
\end{enumerate}

For the ordinary PSH model, \cite{fu} showed the large sample properties of ALASSO, SCAD, and MCP, including selection consistency and the oracle properties, namely, properties of an ideal selection procedure that can work as well as the underlying model were known in advance \citep{scad}. When stratification is added, we can also demonstrate the consistency and the oracle properties of the three penalties under both stratification regimes: when data are regularly stratified, the number of center is finite and the center size goes to infinity; when data are highly stratified, the number of centers goes to infinity whereas the center size holds finite. Theorems stating the large sample properties of $\tbeta^S$ and proofs are included in Appendix A.

%%%%%%%%%%%%%%%%%%%%%%%%%%%%%%%%%%%%%%up to 
\subsubsection{Implementation Issues}
\label{implement s}
The coordinate descent (CD) algorithm \citep{cda, cdacox} and the local quadratic approximation (LQA) algorithm \citep{scad, scadcox} are two implementation methods for penalization methods with convex or non-convex penalties. The former has gained more popularity than the latter for its computational efficiency  \citep{overview}. To penalize the stratified PSH model, we use the LQA algorithm because the CD algorithm may not perform well when the center size is small. 

Two complications arise when applying the CD algorithm to the penalized stratified PSH model with small sizes of centers. First, one step in the CD algorithm is to approximate the objective function by a penalized least square and the optimization problem becomes minimizing $Q^S(\bbeta) \approx n^{-1}\sum_{k=1}^K(\by_k-\boldeta_k)^T(-\partial^2 l_k/\partial \boldeta_k  \partial \boldeta_k^T)(\by_k-\boldeta_k)+\sum_{j=1}^dp_{\lambda}(|\beta_j|)$, where $\boldeta_k=\bZ_k\bbeta$, $\bZ_k$ is the design matrix for stratum $k$ and $\by_k=\boldeta_k-(\partial^2 l_k/\partial \boldeta_k  \partial \boldeta_k^T)^{-1}$ $(\partial l_k/\partial \boldeta_k)$. 
When the center size is small, the event of interest within a center is often rare. Inverting the $n_k\times n_k$ matrix $\partial^2 l_k/\partial \boldeta_k  \partial \boldeta_k^T$ may not be computationally feasible, resulting in the failure of the CD algorithm. For example, three patients in center $k$ have graft failure, DWFG, and loss to follow-up sequentially, then $\partial^2 l_k/\partial \boldeta_k  \partial \boldeta_k^T$ is not positive definite, thus can not be inverted. 
Second, when the inversion is possible, to reduce computational burden, a conventional method in the CD algorithm is to replace the off-diagonal entries of  $\partial^2 l_k/\partial \eta_k  \partial \eta_k^T$  by zeros  \citep{lasso}. The replacement works when $n_k$ is large, because the off-diagonals are much smaller than the diagonals  \citep{gam}. When $n_k$ is moderate to small, the inverted matrix may not be accurate, potentially leading to incorrect selection results.

%Denote $\bZ_k$ as the design matrix for stratum $k$, and let $\boldeta_k=\bZ_k\bbeta$. The CD algorithm approximates $l^S(\bbeta)$ by the Newton-Raphson update and the optimization problem becomes minimizing a penalized least square, $Q^S(\bbeta) \approx n^{-1}\sum_{k=1}^K(\by_k-\boldeta_k)^T(-\partial^2 l_k/\partial \boldeta_k  \partial \boldeta_k^T)(\by_k-\boldeta_k)+\sum_{j=1}^dp_{\lambda}(|\beta_j|)$, where $\by_k=\boldeta_k-(\partial^2 l_k/\partial \boldeta_k  \partial \boldeta_k^T)^{-1}(\partial l_k/\partial \boldeta_k)$. The $n_k\times n_k$ matrix $\partial^2 l_k/\partial \boldeta_k  \partial \boldeta_k^T$ may not be inversed when the center size is small.
%Even if the inversion is possible, the CD algorithm may not select variables correctly. Usually the off-diagonal entries of  $\partial^2 l_k/\partial \eta_k  \partial \eta_k^T$ are replaced by zeros to reduce computation burden \cite{lasso}. The replacement works when $n_k$ is large, because the off-diagonals are much smaller than the diagonals \cite{gam}. When $n_k$ is moderate to small, the inversed matrix may not be accurate, potentially leading to incorrect selection results. 

To implement the LQA procedure, we follow \cite{scad, scadcox} to locally approximate the penalty by a quadratic function. 
Given an initial value $\bbeta^{(0)}$ that is close to $\tbeta^S$, the penalty function $p_{\lambda}(|\beta_j|)$ is quadratically approximated by
\begin {equation*}
p_{\lambda}(|\beta_j|) \approx p_{\lambda}(|\beta_j^{(0)}|) + \dfrac{1}{2}\{p'_{\lambda}(|\beta_j^{(0)}|)/|\beta_j^{(0)}\}\{\beta_j^2-(\beta_j^{(0)})^2\}.
\end{equation*}
Define the score function $U^S(\bbeta)=\partial l^S(\bbeta)/\partial \bbeta$ and the Hessian matrix $H^S(\bbeta)=\partial l^S(\bbeta)/\partial \bbeta \partial \bbeta^T$. We then apply the Newton-Raphson algorithm to maximize the objective function (\ref{objective func s}). At iteration $k+1$, the solution is updated by
\begin{equation}
\label{NR}
\bbeta^{(k+1)}=\bbeta^{(k)}-\{H^S(\bbeta^{(k)})-nA_{\lambda}(\bbeta^{(k)})\}^{-1}\{U^S(\bbeta^{(k)})-n\bbeta^{(k)}A_{\lambda}(\bbeta^{(k)})\},
\end{equation}
where $A_{\lambda}(\bbeta)$ is a diagonal matrix with entires $A_j=\dfrac{p'_{\lambda}(|\beta_j|)}{|\beta_j|}, j=1, \dots, d$. To avoid numerical difficulties when $\beta_j=0$ , we take the strategy from \cite{MM} and employ $A_{j, \epsilon}=\dfrac{p'_{\lambda}(|\beta_j|)}{|\beta_j|+\epsilon}$, where $\epsilon$ is a small positive value such as $10^{-6}$.  From equation (\ref{NR}),  standard errors for non-zero penalized estimators can be directly obtained using the sandwich formula,
\begin{equation*}
\hat{\text{cov}}(\tbeta_1^S)=\{H^S(\tbeta_1^S)-nA_{\lambda}(\tbeta_1^S)\}^{-1}\hat{\text{cov}}\{U^S(\tbeta_1^S)\}\{H^S(\tbeta_1^S)-nA_{\lambda}(\tbeta_1^S)\}^{-1}.
\end{equation*}
The explicit formula of $\hat{\text{cov}}\{U^S(\tbeta_1^S)\}$ can be found in \cite{zhoustratified}. 
%The estimated covariance matrix of the score function is $n\sum_{k=1}^Kp_k\hat\bSigma_{11k}$ for the regularly stratified data and $K\hat\Sigma_{11}$ for highly stratified data. The two formulas are consistent with the results in Theorem 1 and 2. 

We employ the Bayesian Information Criteria (BIC)  as in \cite{bic} to select tuning parameters:
$\BIC(\lambda)=-2l^S(\boldsymbol{\tbeta^S})+\log(n)\hbox{DF}_{\lambda}$, where DF$_{\lambda}$ is the effective number of parameters and can be computed by $\tr[\{H^S(\tbeta^S)-nA_{\lambda}(\tbeta^S)\}^{-1}H^S(\tbeta^S)]$. The optimum $\lambda$ minimizes $\BIC(\lambda)$. For SCAD and MCP, we also need to select values of $\alpha$ and $\gamma$. A two-dimensional grid search is usually conducted to find the best pair of $(\lambda, \alpha)$ or $(\lambda, \gamma)$, but computation can be extensive. \cite{scad} suggested to fix $\alpha \approx 3.7$ for SCAD from a Bayesian statistical point of view. \cite{mcp} suggested to fix $\gamma \approx 2.7$ for MCP. For simplicity, we use their suggested values throughout this paper. 

%%%%%%%%%%%%%%%%%%%%%%%%%%%%%%%%%%%%%%
%%%%%%%%%%%%%%%%%%%%%%%%%%%%%%%%%%%%%%
\subsection{Penalized Marginal PSH Model}
\label{sec2.3}
Different from the penalization strategy in Section (\ref{sec2.2}), which identifies important variables for the cumulative incidence function of graft failure conditional on transplant centers, the primary interest of the penalized marginal PSH model is  the marginal cumulative incidence function. The standard error of regression coefficients is then corrected for within-center correlations. 
The marginal cumulative incidence function is defined as $F_1(t; \bZ_{ki})=P(T_{ki}\leq t, \epsilon_{ki}=1|\bZ_{ki}), i=1, \dots, n_k, k=1, \dots, K$, and the marginal subdistribution hazard can be defined accordingly: $u_1(t; \bZ_{ki})=d F_1(t; \bZ_{ki})/\{1-F_1(t; \bZ_{ki})\}$. \cite{zhoucluster} proposed to model $u_1(t; \bZ_{ki})$ with the proportional hazard specification, $u_1(t; \bZ_{ki})=u_{10}(t)\exp(\bbeta^T\bZ_{ki})$, where $\mu_{10}(t)$ is an unspecified baseline shared by all centers and $\bbeta$ has a population-average interpretation. Relative risk of a factor compares two patients who are randomly selected from the population \citep{LJwei}.  The marginal PSH model can be regarded as an analogy of the marginal Cox model proposed by \cite{lee1992cox} for the competing risks setting, and it is suitable when the scientific interest is on the study population. 

Maximizing the marginal partial likelihood is usually prohibitive in terms of computation. \cite{zhoucluster} proposed a pseudo-partial likelihood function as an alternative to the full likelihood by assuming that event times of graft failure are independent of each other even if they belong to the same center. The log-pseudo-partial likelihood function is defined as follows
\begin{equation}
\label{likelihood_marginal}
\begin{aligned}
l^M(\bbeta)=&\sum_{k=1}^K\sum_{i=1}^{n_k}\delta_{ki}I(\epsilon_{ki}=1)\{\beta^T\bZ_{ki}-\\
&\log \sum_{k'=1}^K\sum_{i'=1}^{n_k}\hw_{k'i'}(X_{k'i'})Y_{k'i'}(X_{k'i'})\exp(\beta^T\bZ_{k'i'})\},
\end{aligned}
\end{equation}
where $\hw_{ki}(t)$ is a marginal inverse probability of censoring weight and is defined as $\hw_{ki}(t)=I(C_{ki}\geq T_{ki}\land t)\hat G(t)/\hat G(X_{ki}\land t), i=1, \dots, n_k, k=1, \dots, K$. Note the formulation of the log-pseudo-partial likelihood function is the same as the log-partial likelihood for the ordinary PSH model. Maximizing this likelihood function is computationally advantageous and the estimated regression coefficients are consistent and asymptotically normal \citep{zhoucluster}. Denote the maximizer of function (\ref{likelihood_marginal}) as $\hbeta^M$. The variance estimator of $\hbeta^M$ is adjusted for the within-center correlations  \citep{zhoucluster}. 

To identify important variables for the marginal cumulative incidence function, we propose to penalized the log-pseudo-partial likelihood function. 
The objective function is defined as 
\begin{equation}
Q^M(\bbeta)=l^M(\bbeta)-K \sum_{j=1}^dp_{\lambda}(|\beta_j|).
\label{objective func m}
\end{equation}
The penalized estimator maximizes the objective function (\ref{objective func m}) and is denoted as 
\begin{equation*}
\tbeta^M=\argmax Q^M(\bbeta).
\end{equation*}
The marginal PSH model regards center as the sampling unit, we hence formulate the objective function (\ref{objective func m}) with a scale of $K$ for the penalty term. With such a formulation, we are able to show selection consistency and the oracle properties of  ALASSO, SCAD, and MCP as $K \to \infty$. The theorem and proofs are included in Appendix B.

The efficient CD algorithm can be employed to optimize the objective function (\ref{objective func m}), because in the penalized marginal PSH model, this algorithm minimizes $Q^M(\bbeta)\approx K^{-1}(\by -\boldeta)^T(-\partial^2l^M/\partial \boldeta \partial \boldeta^T)(\by - \boldeta)+\sum_{j=1}^d p_{\lambda}(|\beta_j|)$, where $\eta=\bZ \bbeta$, $\bZ$ is the $n\times d$ design matrix, and $\by=\boldeta-(\partial^2 l^M/\partial \boldeta  \partial \boldeta^T)^{-1}(\partial l^M/\partial \boldeta)$.
For a data set with a moderate sample size, the $n\times n$ matrix $\partial^2l^M/\partial \eta \partial \eta^T$ is positive definite and thus can be inverted.
Since $Q^M(\bbeta)$ differs from the objective function of the penalized PSH model only by the scale of the penalty term, we follow the implementation procedure presented in  \cite{fu} with adjusted penalty terms. The tuning parameter can be selected by $\BIC(\lambda)=-2l^M(\boldsymbol{\tbeta^M})+\log(K)\hbox{DF}_{\lambda}$, where $\hbox{DF}_{\lambda}= \tr\{\bZ(\bZ^T\hat \bD^M\bZ-K\bA_{\lambda})^{-1}\bZ^T\hat \bD^M\}$ and $\hat\bD^M$ is a diagonal matrix with entries of $h_{ki}=\partial ^2l^M/\partial \eta_{ki}\partial \eta_{ki}^T\arrowvert_{\bbeta=\tbeta^M}$,  $i=1, \dots, n_k, k=1, \dots, K$.

Center effects (or within-center correlations)  are accommodated in the variance-covariance estimator of $\tbeta^M$, which is defined as  
\begin{equation*}
\hat{\text{cov}}(\tbeta^M)=\{\bZ^T\hat \bD^M\bZ-K\bA_{\lambda}(\tbeta^M)\}^{-1}\hat{\text{cov}}(\bU^M(\tbeta^M))\{\bZ^T\hat \bD^M\bZ-K\bA_{\lambda}(\tbeta^M)\}^{-1},
\end{equation*}
where $\bU^M(\bbeta)$ is the score function of $l^M(\bbeta)$ accommodating the within-center correlations. Since we assume event times are independent, $\bU^M(\bbeta)$ is equivalent to the score function of the ordinary PSH model. However, the standard covariance of $\bU^M(\bbeta)$ proposed by \cite{finegray} may no longer be valid due to the possible dependence among patients within a center  \citep{lee1992cox, zhoucluster}. 
\cite{zhoucluster} showed that $\bU^M(\bbeta)$ is asymptotically equivalent to a sum of independent identically distributed random variables. Therefore, the covariance of $\bU^M(\bbeta)$ can be consistently estimated, in a manner robust to the within-cluster correlations. The explicit expression of $\hat{\text{cov}}\{\bU^M(\bbeta)\}$ can be found in \cite{zhoucluster}.

%Without penalization, the robust variance-covariance estimator of $\hbeta^M$ is $\{H^M(\hbeta^M)\}^{-1}\hat{\text{cov}}\{\bU^M(\hbeta^M)\}\{H^M(\hbeta^M)\}^{-1}$, where $H^M(\bbeta)$ is the Hessian matrix of $l^M(\bbeta)$ and $\hat{\text{cov}}\{\bU^M(\bbeta)\}$ is a correction factor that accounts for within-center correlations \cite{zhoucluster}. Similarly, in equation (\ref{Mvar}), $\hat{\text{cov}}\{\bU^M(\bbeta)\}$ accomodates within-center correlations.
 
%%%%%%%%%%%%%%%%%%%%%%%%%%%%%%%%%%%%%%%%%%%%%%%%%%%%%
\subsection{Extension to Group Variable Selection}
\label{sec 2.4}
Among the potential risk factors for graft failure, some have multiple categories and can be represented by a group of dummy variables. For example, the mismatch score of human leukocyte antigen B (HLA-B) has three levels (0, 1, and 2) and the peak panel reactive antibody is categorized into four levels (=0, 1-50, 51-80, $>$80). The selection of such factors corresponds to the selection of groups of variables. Our methods can be easily adapted to group variable selection by adjusting the penalty function.

Assume there are $J$ groups of variables and each group is of size $d_j, j=1, \dots, J$. The vector of regression coefficients can be divided into $J$ sub-vectors, $\bbeta^T=(\bbeta_1^T, \dots, \bbeta_J^T)$. We replace the penalty functions in objective functions (\ref{objective func s}, \ref{objective func m}) with their group version, $p(\|\bbeta_j\|; d_j^{1/2}\lambda)$, where $\|\bbeta_j\|=(\bbeta_j^T\bbeta_j)^{1/2}$ and $d_j$ is a scalar for adjusting group sizes, $j=1, \dots, J$. The case $d_1 = \dots, = d_J=1$ denotes individual variable selection. Following \cite{fu}, the penalized estimators with group penalties are expected to behave similarly to their individual versions and optimization procedures can be developed accordingly.
%%%%%%%%%%%%%%%%%%%%%%%%%%%%%%%%%

%%%%%%%%%%%%%%%%%%%%%%%%%%%%%%%%%
\section{Simulation}
\label{sec3}
Three sets of simulations are conducted to assess the performance of the penalized stratified PSH model (`penalized stratified model' hereafter) and the penalized marginal PSH model (`penalized marginal model' hereafter). 
Results on the maximum partial likelihood estimator (MPLE) and the oracle estimators are also reported. Here  the oracle estimator is obtained by fitting the underlying model. In the first simulation, the number of centers is fixed at three and the center size is relatively large. The performance of the penalized stratified model for regular stratification is evaluated.  The penalized marginal method is not assessed due to the small number of centers. In the second set, the center number is large while the center size is small. The performance of the penalized stratified model for high stratification and the penalized marginal model is evaluated. In the third set, both the number of centers and the center size are moderate. The performance of the penalized stratified model for regular and high stratification as well as the penalized marginal model is assessed.
%are not assessed because they regards a center as a unit and the selection consistency and the oracle properties can only be achieved when the center number is large.
 
In the first two sets of simulations, we also applied the penalized PSH model with log-normal frailties (`penalized frailty model' hereafter)  \citep{haselection} and compare its performance with the penalized stratified model. The penalized frailty model is a conditional approach that models the center effect through frailties. For center $k$, the subdistribution hazard conditional on frailties $\bv_1=(v_{11}, \dots, v_{1k}, \dots, v_{1K})$ has the following form
\begin{equation}
\label{frailty}
\lambda_{1k}(t; \bZ_{ki}, v_{1k})=\lambda_{10}(t)v_{1k}\exp(\bbeta_1^T\bZ_{ki}),
\end{equation}
where $\bv_1$ has two assumptions: (1) $\bv_1$ follows a log-normal distribution and (2) $\bv_1$ acts proportionally on the baseline subdistribution hazard. Through penalizing the model in equation (\ref{frailty}), \cite{haselection} demonstrated that the model performs well through simulations where both assumptions on $\bv_1$ are satisfied. In our simulations, we show the limitation of the penalized frailty model due to its restrictive assumptions, and also demonstrate the flexibility of the penalized stratified model, which does not require explicit modeling of the center effect.

The performance of variable selection methods is assessed by four measures: the average numbers of correctly (C) and incorrectly (IC) selected zero coefficients, the percentage of identifying the true model (Pcorr), and the median of mean squared error (MMSE). Values of C and IC characterize the performance in shrinking unimportant variables to zero and selecting important variables respectively. The mean squared error (MSE) is defined as  $(\tbeta^S-\bbeta_0)\boldsymbol{I}(\tbeta^S-\bbeta_0)$ or $(\tbeta^M-\bbeta_0)\boldsymbol{I}(\tbeta^M-\bbeta_0)$ for assessing the model error, where $\boldsymbol{I}$ is the population correlation matrix  \citep{lassocox} .

\subsection{A Small Number of Large Centers} 
\label{sim1}
In this scenario, the center indicator $\xi$ is randomly sampled from $\{1, 2, 3\}$. For each center, we follow the procedure described in \cite{finegray} to simulate data. Eight covariates $Z_1, \dots, Z_8$ are included in the model, and they are marginally standard normal with pairwise correlations corr$(Z_i, Z_j)=\rho^{|i-j|}$, where $\rho=0.5$. For cause 1, the vector of regression coefficients is $\bbeta_1=(0.8, 0, 0, 1, 0, 0, 0.6, 0)^T$, implying that the underlying model includes $Z_1, Z_4$, and $Z_7$. For cause 2, the regression parameter is $\bbeta_2= -\bbeta_1$. The baseline subdistribution hazards for the three centers have different parametric forms and their cumulative incidence functions for cause 1 are defined as follows
\begin{align*}
F_{11}(t; \bZ_{1i})&=1-\{1-p\times \Phi(\dfrac{\log(t)-1}{0.25})\}^{\exp(\bbeta_1^T\bZ_{1i})}, \\
F_{21}(t; \bZ_{2i})&=1-[1-p\{1-\exp(-0.018e^t+0.018)\}]^{\exp(\bbeta_1^T\bZ_{2i})}, \\
F_{31}(t; \bZ_{3i})&=1-[1-p\{1-\exp(-t^5)\}]^{\exp(\bbeta_1^T\bZ_{3i})},
\end{align*}
which are a log-normal mixture, a Gompertz mixture, and a Weibull mixture with mass $1-p$ at $\infty$ when $\bZ_{ki}=0$. The value of $p$ is $0.6$. Their corresponding cumulative incidence functions conditional on cause 2 follow exponential distributions with rates $5\exp(\bbeta_2^T\bZ_{1i})$, $10\exp(\bbeta_2^T\bZ_{2i})$, and $2\exp(\bbeta_2^T\bZ_{3i})\}$. Censoring times are independently generated from a Uniform (0, 9) distribution to achieve a censoring rate of $25\% - 30\%$. Note under this setting, the penalized frailty model may not perform well due to the restrictive form of the subdistribution hazard. According to the model in equation (\ref{frailty}), baseline subdistribution hazards of the three centers only differ by the scalar $v_{1k}$, while the ones in our scenario have distinct shapes. 

Table 1 shows the selection results of the penalization methods for the regularly stratified PSH model and the PSH frailty model with sample sizes of 200 and 400. 
All penalties outperform the MPLE. In particular, the performance of ALASSO, SCAD, and MCP in identifying the underlying model and parameter estimation is close to the oracle estimator. LASSO tends to select over-fitted models with higher false positive rates than the other three penalties. Regardless of penalty choices, the penalized frailty model tends to select unimportant variables while dropping the important ones. Its model error is twice the error of the penalized stratified model when $n=200$ and five times the error when $n=400$.

\begin{table}[t]
\centering
\begin{threeparttable}
\caption{Selection results of the penalized stratified and frailty model based on 100 replications with $p=0.6$. Center number $K=3$. Sample size $n=200$ and $400$. The censoring rate is 28\% and the event rate is 40\%.}
\label{table 1}
\begin{tabular}{llllrlllrl}
\hline\noalign{\smallskip}
       &            & \multicolumn{4}{c}{n=200}   & \multicolumn{4}{c}{n=400}      \\
Penalty&   Model  & C    & IC   & Pcorr & MMSE  & C    & IC & Pcorr & MMSE   \\
\noalign{\smallskip}\hline\noalign{\smallskip}
MPLE   & Stratified & 0    & 0    & 0\%   & 0.133& 0    & 0 & 0\%   & 0.060  \\
     & Frailty    & 0    & 0    & 0\%   & 0.280 & 0    & 0 & 0\%   & 0.219  \\
LASSO  & Stratified & 3.37 & 0    & 23\%  & 0.112 & 3.76 & 0  & 29\%  & 0.055  \\
       & Frailty    & 3.50  & 0.05 & 23\%  & 0.228& 3.78 & 0  & 32\%  & 0.180 \\
ALASSO & Stratified & 4.75 & 0.01 & 79\%  & 0.064  & 4.92 & 0  & 92\%  & 0.028\\
       & Frailty    & 4.57 & 0.08 & 69\%  & 0.159  & 4.77 & 0  & 83\%  & 0.118  \\
SCAD   & Stratified & 4.92 & 0    & 92\%  & 0.052 & 4.93 & 0  & 93\%  & 0.027  \\
       & Frailty    & 4.74 & 0.12 & 81\%  & 0.100 & 4.85 & 0  & 86\%  & 0.100  \\
MCP    & Stratified & 4.92 & 0    & 92\%  & 0.056  & 4.93 & 0  & 93\%  & 0.026\\
       & Frailty    & 4.76 & 0.12 & 84\%  & 0.098 & 4.85 & 0  & 86\%  & 0.100  \\
Oracle & Stratified & 5    & 0    & 100\% & 0.044  & 5    & 0  & 100\% & 0.021\\
       & Frailty    & 5    & 0    & 100\% & 0.095 & 5    & 0  & 100\% & 0.092 \\
\noalign{\smallskip}\hline
\end{tabular}
\begin{tablenotes}
\footnotesize
\item C: the average number of coefficients that are correctly set to zero; IC: the average number of coefficients that are incorrectly set to zero; Pcorr: the percentage of identifying the true model; MMSE: median of mean squared error.
\end{tablenotes}
\end{threeparttable}
\end{table}

\subsection{A Large Number of Small Centers} 
\label{sim2}
In this scenario, center numbers are 100 and 200. The center size is either 2 or uniformly sampled from $\{2, 3, 4, 5\}$. For center $k$, we consider a conditional subdistribution hazard in the form of the frailty model, $\lambda_{1k}(t; \bZ_{ki}, v_{1k})=\lambda_{10}(t)v_{1k}\exp(\bbeta_1^T\bZ_{ki})$, where $\lambda_{10}(t)=e^{-t}$ and $v_{1k}$ is generated from a positive stable distribution with parameter $\alpha_1$.  A smaller $\alpha_1$ is associated with a higher within-center correlation, implying that patients within a center are more similar. Under this setting, the assumption that $v_{1k}$ follows a log-normal distribution in the penalized frailty model is violated.
Covariates $Z_1, \dots, Z_8$ are marginally standard normal with pairwise correlations $\rho^{|i-j|}$, where $\rho=0.5$. The cumulative incidence function conditional on frailties and covariates is $F_{1k}(t; \bZ_{ki}, v_{1k})=1-\exp\{e^{-t}v_{1k}\exp(\bbeta_1^T\bZ_{ki})\}$. 
Given cause 2, the conditional cumulative incidence function is assumed to be exponential with a rate of $v_{2k}\exp(\bbeta_2^T\bZ_{ki})$, where $\bbeta_2=-\bbeta_1$. The frailty $v_{2k}$ is generated from a positive stable distribution with parameter $\alpha_2$. Data are then simulated based on the two conditional cumulative incidence functions.

From the conditional cumulative incidence function, we can derive the marginal cumulative incidence function and the marginal subdistribution hazard by integrating out the frailties. 
The resulting marginal subdistribution hazard also satisfies the proportional hazard assumption  \citep{zhoucluster, marginallogan}. By Laplace transformation,
\begin{align}
F_1(t; \bZ_{ki})&=\int_0^tF_1(u; \bZ_{ki}, v_{1k})dF_{V_{1k}}(u)=1-\exp\{-M_0^{\alpha_1}(t)e^{(\bbeta_1^*)^T\bZ_{ki}}\},\\
\label{connect}
u_{1}(t; \bZ_{ki})&=\alpha_1 M_0^{\alpha_1-1}(t)\lambda_{10}(t)\exp\{(\bbeta_1^*)^T\bZ_{ki}\},
\end{align} 
where $M_0(t)=\int_0^t\lambda_{10}(s)ds$, and $\bbeta_1^*=\alpha_1 \bbeta_1$.  Correspondingly, the marginal cumulative incidence function given cause 2 is exponential with rate $\alpha_2 t^{\alpha_2-1} \exp\{(\bbeta_2^*)^T\bZ_{ki}\}$ and $\bbeta_2^*=\alpha_2\bbeta_2$.

Table 2 summarizes the selection results of the penalization method for the highly stratified PSH model and the PSH frailty model, where  $\bbeta_1=(0.8, 0, 0, 1, 0, 0, 0.6, 0)$ and $\alpha_1=\alpha_2=0.4$ or $0.7$. The penalized frailty model is only applied for $\alpha_1=0.7$, because when $\alpha_1=0.4$, the algorithm can not converge under the mis-specification of the frailty distribution.  We additionally include relative MMSE, which is defined as the ratio of the MMSE of the penalized estimator to the oracle estimator. 

\begin{table}[t]
\centering
\caption{Selection results of the penalized stratified and frailty model based on 100 replications for $\alpha_1=0.4$ and $0.7$. Center number $K=100$ and $200$. Center size $n_k=2$ and $n_k \in \{2, 3, 4, 5\}$. The censoring rate is 27\% and the event rate is 46\%. Numbers in the parentheses are ratios of MMSE of the penalized estimator to the oracle estimator.}
\label{table 2}
\resizebox{\textwidth}{!}{
\begin{threeparttable}
\begin{tabular}{lllllllllllllll}
\hline\noalign{\smallskip}
&    &         & \multicolumn{8}{c}{Stratified} & \multicolumn{4}{c}{Frailty}     \\
&    &         & \multicolumn{4}{c}{$\alpha_1=0.4$}     & \multicolumn{4}{c}{$\alpha_1=0.7$}     & \multicolumn{4}{c}{$\alpha=0.7$}            \\ 
\noalign{\smallskip}\hline\noalign{\smallskip}
$n_k$ & $K$  & penalty & C    & IC   & Pcorr & MMSE     & C    & IC   & Pcorr & MMSE(ratio)        & C    & IC & Pcorr & MMSE(ratio)        \\
\noalign{\smallskip}\hline\noalign{\smallskip}
2 & 100 & MPLE & 0 & 0 & 0\% & 1.184 & 0 & 0 & 0\% & 1.160 (2.86) & 0 & 0 & 0\% & 0.176 (2.63) \\
 &  & LASSO & 3.93 & 0.41 & 29\% & 0.616 & 3.76 & 0.38 & 30\% & 0.557 (1.37) & 3.12 & 0.02 & 15\% & 0.195 (2.92) \\
 &  & ALASSO & 4.45 & 0.61 & 46\% & 0.494 & 4.42 & 0.50 & 47\% & 0.541 (1.33) & 4.27 & 0.06 & 49\% & 0.125 (1.87) \\
 &  & SCAD & 4.36 & 0.81 & 46\% & 0.940 & 4.53 & 0.81 & 43\% & 0.964 (2.37) & 4.60 & 0.09 & 69\% & 0.093 (1.40) \\
 &  & MCP & 4.19 & 0.68 & 44\% & 0.986 & 4.52 & 0.80 & 42\% & 0.964 (2.37) & 4.60 & 0.08 & 69\% & 0.102 (1.52) \\
 &  & Oracle & 5 & 0 & 100\% & 0.283 & 5 & 0 & 100\% & 0.406 (1.00) & 5 & 0 & 100\% & 0.067 (1.00) \\
 & 200 & MPLE & 0 & 0 & 0\% & 0.365 & 0 & 0 & 0\% & 0.406 (3.09) & 0 & 0 & 0\% & 0.098 (2.18) \\
 &  & LASSO & 3.58 & 0.08 & 22\% & 0.307 & 3.54 & 0.03 & 17\% & 0.320 (2.44) & 3.34 & 0 & 20\% & 0.116 (2.58) \\
 &  & ALASSO & 4.60 & 0.18 & 57\% & 0.226 & 4.58 & 0.18 & 59\% & 0.241 (1.84) & 4.69 & 0 & 76\% & 0.075 (1.67) \\
 &  & SCAD & 4.79 & 0.35 & 71\% & 0.218 & 4.83 & 0.34 & 69\% & 0.287 (2.18) & 4.76 & 0 & 81\% & 0.060 (1.34) \\
 &  & MCP & 4.79 & 0.36 & 72\% & 0.218 & 4.82 & 0.33 & 69\% & 0.280 (2.13) & 4.76 & 0 & 81\% & 0.060 (1.34) \\
 &  & Oracle & 5 & 0 & 100\% & 0.102 & 5 & 0 & 100\% & 0.131 (1.00) & 5 & 0 & 100\% & 0.045 (1.00) \\
\noalign{\smallskip}\hline\noalign{\smallskip}
\{2,3,4,5\} & 100 & MPLE & 0 & 0 & 0\% & 0.169 & 0 & 0 & 0\% & 0.164 (2.80) & 0 & 0 & 0\% & 0.085  (2.75) \\
 &  & LASSO & 3.66 & 0 & 20\% & 0.161 & 3.61 & 0 & 22\% & 0.178 (3.04) & 3.34 & 0 & 19\% & 0.100 (3.23) \\
 &  & ALASSO & 4.70 & 0.05 & 72\% & 0.104 & 4.76 & 0.02 & 79\% & 0.097 (1.65) & 4.58 & 0 & 68\% & 0.058 (1.89) \\
 &  & SCAD & 4.86 & 0.06 & 86\% & 0.060 & 4.90 & 0.06 & 88\% & 0.060 (1.02) & 4.70 & 0 & 75\% & 0.051 (1.63) \\
 &  & MCP & 4.87 & 0.06 & 87\% & 0.060 & 4.90 & 0.05 & 87\% & 0.060 (1.02) & 4.70 & 0 & 75\% & 0.051 (1.63) \\
 &  & Oracle & 5 & 0 & 100\% & 0.049 & 5 & 0 & 100\% & 0.058 (1.00) & 5 & 0 & 100\% & 0.031 (1.00) \\
 & 200 & MPLE & 0 & 0 & 0\% & 0.075 & 0 & 0 & 0\% & 0.078 (2.47) & 0 & 0 & 0\% & 0.042 (2.45) \\
 &  & LASSO & 3.49 & 0 & 23\% & 0.087 & 3.76 & 0 & 29\% & 0.080 (2.55) & 3.44 & 0 & 18\% & 0.047 (2.77) \\
 &  & ALASSO & 4.85 & 0 & 87\% & 0.038 & 4.82 & 0 & 84\% & 0.041 (1.29) & 4.72 & 0 & 79\% & 0.024 (1.42) \\
 &  & SCAD & 4.94 & 0 & 94\% & 0.030 & 4.92 & 0 & 93\% & 0.034 (1.08) & 4.76 & 0 & 79\% & 0.022 (1.31) \\
 &  & MCP & 4.94 & 0 & 94\% & 0.030 & 4.92 & 0 & 93\% & 0.034 (1.08) & 4.78 & 0 & 80\% & 0.022 (1.27) \\
 &  & Oracle & 5 & 0 & 100\% & 0.028 & 5 & 0 & 100\% & 0.032 (1.00) & 5 & 0 & 100\% & 0.017 (1.00)\\
\noalign{\smallskip}\hline
\end{tabular}
\begin{tablenotes}
\footnotesize
\item  Notation is the same as Table \ref{table 1}.
  \end{tablenotes}
 \end{threeparttable}}
\end{table}

%The center effect is absorbed into the unspecified $\lambda_{1k0}(t)$ and thus it is not explicitly modeled. 

Overall, the performance of the penalized stratified model is improved as the sample size increases, but invariant to the change of within-center correlations. 
When the center size is 2 such as in the matched cohort study,  error rates of identifying important variables and model errors are much higher than the oracle estimator. Because a center only contributes to the partial likelihood when the failure from the cause of interest is observed for a subject  while the other subject has not failed from any causes or has already failed from the competing risks, a great amount of information is discarded, leading to the worsened performance. 
As the center size increases to $\{2, 3, 4, 5\}$, the accuracy in model selection and parameter estimation is considerably improved. SCAD and MCP consistently beat the other penalties and perform similarly to the oracle estimator in all scenarios. 

Comparing the penalized stratified and the penalized frailty models for $\alpha_1=0.7$, the former outperforms the latter in eliminating unimportant variables and producing small relative MMSEs when the center size is in $\{2, 3, 4, 5\}$, but underperforms when the center size is two. 
Note that the penalized frailty model has smaller absolute MMSEs than the penalized stratified model for all penalties, MPLE, and the oracle estimator.  It is possible that the parametric modeling of frailties using the log-normal distribution can roughly capture the positive stable distribution, leading to less variation in penalized estimators than the penalized stratified model.
%It is possible that the log-normal distribution of frailties can roughly capture the positive stable distribution. Thus, estimated regression coefficients are less variant and model errors are smaller than the penalized stratified model.

\begin{table}[t]
\centering
\caption{Selection results of the penalized marginal model based on 100 replications with $\alpha_1=0.4$ and $0.7$. Center number $K=100$ and $200$. Center size $n_k=2$ and $n_k \in \{2, 3, 4, 5\}$. The censoring rate is 29\% and the event rate is 43\%.}
\label{table 3}
\scalebox{0.85}{
\begin{threeparttable}
\begin{tabular}{lllllllllll}
\hline\noalign{\smallskip}
 &  &  & \multicolumn{4}{l}{$\alpha_1=0.4$} & \multicolumn{4}{l}{$\alpha_1=0.7$} \\
$n_k$ & $K$ & penalty & C & IC & Pcorr & MMSE & C & IC & Pcorr & MMSE \\
\noalign{\smallskip}\hline\noalign{\smallskip}
2         & 100 & MPLE    & 0            & 0  & 0\%   & 0.166 & 0            & 0  & 0\%   & 0.171 \\
          &     & LASSO   & 3.55         & 0  & 26\%  & 0.092 & 3.39         & 0  & 19\%  & 0.130 \\
          &     & ALASSO  & 4.73         & 0  & 77\%  & 0.048 & 4.74         & 0  & 79\%  & 0.064 \\
          &     & SCAD    & 4.79        & 0  & 82\%  & 0.040 & 4.83         & 0  & 86\%  & 0.054 \\
          &     & MCP     & 4.83         & 0  & 85\%  & 0.039 & 4.80          & 0  & 82\%  & 0.054 \\
          &     & Oracle  & 5            & 0  & 100\% & 0.030 & 5            & 0  & 100\% & 0.044 \\
          & 200 & MPLE    & 0            & 0  & 0\%   & 0.076 & 0            & 0  & 0\%   & 0.066 \\
          &     & LASSO   & 3.29         & 0  & 19\%  & 0.063 & 3.50          & 0  & 20\%  & 0.069 \\
          &     & ALASSO  & 4.89         & 0  & 89\%  & 0.026 & 4.88         & 0  & 90\%  & 0.023 \\
          &     & SCAD    & 4.90         & 0  & 92\%  & 0.021 & 4.92         & 0  & 93\%  & 0.017 \\
          &     & MCP     & 4.91          & 0  & 92\%  & 0.022 & 4.89         & 0  & 90\%  & 0.018 \\
          &     & Oracle  & 5            & 0  & 100\% & 0.020 & 5            & 0  & 100\% & 0.016 \\
\noalign{\smallskip}\hline\noalign{\smallskip}
(2,3,4,5) & 100 & MPLE   & 0    & 0 & 0\%   & 0.099 & 0    & 0 & 0\%   & 0.097 \\
          &     & LASSO  & 3.38 & 0 & 20\%  & 0.066 & 3.56 & 0 & 22\%  & 0.060 \\
          &     & ALASSO & 4.81 & 0 & 87\%  & 0.032 & 4.84 & 0 & 85\%  & 0.035 \\
          &     & SCAD   & 4.80  & 0 & 87\%  & 0.031 & 4.88 & 0 & 91\%  & 0.031 \\
          &     & MCP    & 4.81 & 0 & 87\%  & 0.030 & 4.84 & 0 & 85\%  & 0.032 \\
          &     & Oracle & 5    & 0 & 100\% & 0.028 & 5    & 0 & 100\% & 0.029 \\
          & 200 & MPLE   & 0    & 0 & 0\%   & 0.050 & 0    & 0 & 0\%   & 0.039 \\
          &     & LASSO  & 3.48 & 0 & 18\%  & 0.036 & 3.73 & 0 & 28\%  & 0.032 \\
          &     & ALASSO & 4.93 & 0 & 93\%  & 0.017 & 4.95 & 0 & 95\%  & 0.016 \\
          &     & SCAD   & 4.97 & 0 & 97\%  & 0.015 & 4.96 & 0 & 96\%  & 0.014 \\
          &     & MCP    & 4.95 & 0 & 95\%  & 0.016 & 4.96 & 0 & 96\%  & 0.014 \\
          &     & Oracle & 5    & 0 & 100\% & 0.014 & 5    & 0 & 100\% & 0.014\\
\noalign{\smallskip}\hline
\end{tabular}
\begin{tablenotes}
\footnotesize
\item  Notation is the same as Table \ref{table 1}.
  \end{tablenotes}
 \end{threeparttable}}
\end{table}
 
Table 3 shows the simulation results for the penalized marginal PSH model with $\bbeta^*_1=(0.8, 0, 0, 1, 0, 0, 0.6, 0)^T$ and $\alpha_1=\alpha_2=0.4$ or $0.7$. For both center sizes, the performance is similar. As the number of centers increases, the accuracy of variable selection and parameter estimation is improved.  Three penalties, ALASSO, SCAD, and MCP have close performance to the oracle estimator for $K=200$. When the within-center correlation is larger ($\alpha_1=0.4$), selected models have less correctly selected zeros and higher model errors versus when the correlation is smaller ($\alpha_1=0.7$), but the difference is minor.

It should be noted that the penalized stratified model and the penalized marginal model were applied to data that were simulated using different regression parameters (\textit{ i.e.}, $\bbeta_1$). 
In the simulations for the penalized stratified model, data were simulated based on the center-specific subdistribution hazard $\lambda_{1k}(t; \bz_{ki}, v_{1k})=\lambda_{10}(t)v_{1k}\exp(\bbeta_1^T\bz_{ki})$ with $\bbeta_1=(0.8, 0,0,1,0,0,0.6)^T$. 
In the simulations for the penalized marginal model, we set $\bbeta^*_1=(0.8, 0,0,1,0,0,0.6)^T$, and data were generated using the center-specific subdistribution hazard with $\bbeta_1=\alpha_1^{-1}\bbeta^*_1$. Since  $0<\alpha_1<1$, $\bbeta_1$ is larger than $\bbeta^*_1$ in magnitude, implying that the covariate effects for the underlying marginal PSH model is smaller than those for the underlying stratified PSH model. When the penalized stratified model and the penalized marginal model are applied to the same data set,  the latter is more likely to select a larger model than the former, because the penalization method tends to select a larger model when the covariate effects are smaller  \citep{lassocox, alassocox}.

\subsection{A Moderate Number of Moderately Sized Centers} 
\label{sim3}
We follow Section \ref{sim2} to simulate competing risks data with 50 centers and a center size of 25 or 50. Table \ref{table 4} summarizes the selection results of the penalized stratified models with both regular and high stratification, where $\bbeta_1=(0.8, 0, 0, 1, 0, 0, 0.6, 0)$, $\alpha_1=\alpha_2=0.4$ or $0.7$. When both the number of center and the center size are moderate, the performance of the penalized regularly stratified model is similar to that of the penalized highly stratified model. Similar to the observations in Table \ref{table 2}, the within-center correlation does not affect the performance of the penalized stratified model. Three penalties, ALASSO, SCAD, and MCP, perform similarly to the oracle estimator for both $n_k=25$ and 50, due to the large sample size. Notably, the performance of ALASSO for the penalized highly stratified model is  improved compared to its performance when the center size is small in Section \ref{sim2}. Because a larger  center contributes more information, the MPLE,\textit{ i.e.} the data-adaptive weights for the ALASSO, is more accurate. 

 Table \ref{table 4} also demonstrates the selection results for the penalized marginal model with the same settings of center sizes, center numbers, and within-center correlations. Data were simulated with $\bbeta^*_1=(0.8, 0, 0, 1, 0, 0, 0.6, 0)^T$ as in Section \ref{sim2}. When the number of centers is moderate ($K=50$), the penalized marginal model performs worse than the scenario when $K=100$ or $200$ in  Section \ref{sim2}. 
 
In the above simulations in Section \ref{sim1}-\ref{sim3}, censoring times were generated independently from event times, failure causes, and covariates. When censoring times are covariate-dependent, the simulation results suggest that the proposed penalization methods perform similarly to the situation when censoring times and covariates are independent.  This may be due to the inverse probability of censoring weighting techniques  \citep{robins1992recovery} used in the stratified PSH model  \citep{zhoustratified}, and the marginal inverse probability of censoring weight applied in the marginal PSH model  \citep{zhoucluster}, which allows possible dependence between covariates and censoring times  \citep{finegray}. Refer to Appendix D.
 
\begin{table}[h]
\centering
\caption{Selection results of the penalized stratified model based on 100 replications for $\alpha_1=0.4$ and $0.7$. Center number $K=50$. Center size $n_k=25$ and $50$. The censoring rate is 26-27\% and the event rate is 45-49\%.}
\label{table 4}
\scalebox{0.75}{
\begin{threeparttable}
\begin{tabular}{llllllllllll}
\hline\noalign{\smallskip}
   &    &    &        & \multicolumn{4}{c}{$\alpha=0.4$} & \multicolumn{4}{c}{$\alpha=0.7$} \\
Model &  K  & $n_k$    &        & C     & IC  & Pcorr  & MMSE   & C     & IC  & Pcorr  & MMSE   \\
\noalign{\smallskip}\hline\noalign{\smallskip}
Regularly     &50 &25 & MPLE   & 0     & 0   & 0      & 0.018  & 0     & 0   & 0      & 0.017  \\
Stratified   &  &  & LASSO  & 3.56  & 0   & 0.2    & 0.020  & 3.67  & 0   & 0.22   & 0.022  \\
               &    &    & ALASSO & 4.96  & 0   & 0.96   & 0.007  & 4.93  & 0   & 0.94   & 0.018  \\
               &    &    & SCAD   & 4.97  & 0   & 0.97   & 0.005  & 4.97  & 0   & 0.97   & 0.007  \\
               &    &    & MCP    & 4.97  & 0   & 0.97   & 0.005  & 4.97  & 0   & 0.97   & 0.007  \\
               &    &    & Oracle & 5     & 0   & 1      & 0.005  & 5     & 0   & 1      & 0.007  \\
               &    & 50 & MPLE   & 0     & 0   & 0      & 0.008  & 0     & 0   & 0      & 0.006  \\
               &    &    & LASSO  & 3.81  & 0   & 0.2    & 0.009  & 3.77  & 0   & 0.33   & 0.010  \\
               &    &    & ALASSO & 4.95  & 0   & 0.95   & 0.003  & 4.97  & 0   & 0.97   & 0.004  \\
               &    &    & SCAD   & 4.98  & 0   & 0.98   & 0.003  & 4.94  & 0   & 0.94   & 0.004  \\
               &    &    & MCP    & 4.98  & 0   & 0.98   & 0.003  & 4.94  & 0   & 0.94   & 0.004  \\
               &    &    & Oracle & 5     & 0   & 1      & 0.003  & 5     & 0   & 1      & 0.004  \\
               \noalign{\smallskip}\hline\noalign{\smallskip}
Highly     & 50  & 25 & MPLE   & 0     & 0   & 0      & 0.018  & 0     & 0   & 0      & 0.017  \\
Stratified  &  &  & LASSO  & 3.57  & 0   & 0.21   & 0.021  & 3.67  & 0   & 0.23   & 0.023  \\
               &    &    & ALASSO & 4.96  & 0   & 0.96   & 0.007  & 4.92  & 0   & 0.93   & 0.019  \\
               &    &    & SCAD   & 4.97  & 0   & 0.97   & 0.005  & 4.97  & 0   & 0.97   & 0.007  \\
               &    &    & MCP    & 4.97  & 0   & 0.97   & 0.005  & 4.97  & 0   & 0.97   & 0.007  \\
               &    &    & Oracle & 5     & 0   & 1      & 0.005  & 5     & 0   & 1      & 0.007  \\
               &    & 50 & MPLE   & 0     & 0   & 0      & 0.008  & 0     & 0   & 0      & 0.006  \\
               &    &    & LASSO  & 3.78  & 0   & 0.18   & 0.010  & 3.76  & 0   & 0.35   & 0.010  \\
               &    &    & ALASSO & 4.95  & 0   & 0.95   & 0.003  & 4.97  & 0   & 0.97   & 0.004  \\
               &    &    & SCAD   & 4.98  & 0   & 0.98   & 0.003  & 4.94  & 0   & 0.94   & 0.004  \\
               &    &    & MCP    & 4.98  & 0   & 0.98   & 0.003  & 4.94  & 0   & 0.94   & 0.004  \\
               &    &    & Oracle & 5     & 0   & 1      & 0.003  & 5     & 0   & 1      & 0.004 \\
               \noalign{\smallskip}\hline\noalign{\smallskip}
Marginal & 50 & 25 & MPLE   & 0    & 0         & 0     & 0.033 & 0    & 0  & 0     & 0.026 \\
 &  &  & LASSO  & 2.52 & 0         & 0     & 0.025 & 3.04 & 0  & 0.13  & 0.023 \\
          & &    & ALASSO & 4.76 & 0         & 0.81  & 0.017 & 4.86 & 0  & 0.87  & 0.011 \\
          & &    & SCAD   & 4.70  & 0         & 0.83  & 0.016 & 4.71 & 0  & 0.85  & 0.013 \\
          & &    & MCP    & 4.75 & 0         & 0.83  & 0.018 & 4.75 & 0  & 0.84  & 0.013 \\
          & &    & Oracle & 5    & 0         & 1     & 0.014 & 5    & 0  & 1     & 0.010 \\
          & & 50 & MPLE   & 0    & 0         & 0     & 0.025 & 0    & 0  & 0     &   0.018    \\
          & &    & LASSO  & 2.60  & 0         & 0     & 0.018 & 3.39 & 0  & 0.15  & 0.014 \\
          & &    & ALASSO & 4.77 & 0         & 0.82  & 0.016 & 4.85 & 0  & 0.88  & 0.010 \\
          & &    & SCAD   & 4.69 & 0         & 0.84  & 0.016 & 4.77 & 0  & 0.88  & 0.012 \\
          & &    & MCP    & 4.76 & 0         & 0.84  & 0.017 & 4.85& 0   & 0.88      & 0.010      \\
          & &    & Oracle & 5    & 0         & 1     & 0.013 & 5    & 0  & 1     &   0.008\\
                         \noalign{\smallskip}\hline   
\end{tabular}
\begin{tablenotes}
\footnotesize
\item  Notation is the same as Table \ref{table 1}.
  \end{tablenotes}
 \end{threeparttable}}
\end{table}

\section{Application}
\label{sec4}
In the KDRI study,  a subset of UNOS data was used \citep{rao}, including patients who received deceased donor kidney transplants in 1995-2005 with the following exclusion criteria: recipients with previous transplants, with multi-organ transplants, ABO-incompatible, with missing/invalid donor height, with missing/invalid donor weight, or with missing/invalid donor creatinine.  
Patients were followed  from the time of transplantation until the earliest onset of graft failure, DWFG, loss to follow-up, or the conclusion of the observation  period (May 1, 2006). A stratified Cox model adjusting for the transplant center effect was fitted and a composite endpoint including graft failure and DWFG was fitted. 

We use the UNOS data maintaining the same exclusion criteria, but with longer follow-up time. Patients were followed through March 21, 2013, which incorporates more follow-up information than the KDRI study.
We only include patients with complete information on relevant variables.  A total of 33,690 patients from 256 centers were obtained, among whom, 10,357 (31\%) had graft failures,  7,936 (24\%) died with a functioning graft, and 15,397 (46\%) were censored. The median number of patients per center is 98. 

Thirty-four potential risk factors from the KDRI study are considered in the initial model. Donor factors include donor age, race, sex, height, weight, cause of death, donation after cardiac death, serum creatinine, diabetes status, hypertension status, cigarette users, hepatitis C virus (HCV) positivity. Transplant factors include cold ischemia time, organ sharing (local, regional, national), human leukocyte antigen B (HLA-B) mismatch score (0, 1, 2), HLA-DR mismatch score (0, 1, 2), {\it en bloc} transplant, double transplant, and ABO compatibility. Recipient factors include recipient age, race, sex, height, weight, primary diagnosis (glomerulonephritis, diabetes, hypertension, failed transplants, congenital anomalies of the kidney and urinary tract (CAKUT), others), diabetes status, pre-transplant blood transfusion, peak panel reactive antibody (PRA) level (=0, 1-50, 51-80, $>$80), years of renal replacement therapy (RRT) ($\leq$1, 2-3, $>$3), angina pectoris, peripheral vascular disease (PVD), drug-treated chronic obstructive pulmonary disease (COPD), and HCV positivity. Two continuous variables, donor's age and serum creatinine, have nonlinear relationship with the outcome \citep{rao}, we categorize them into three and two levels respectively as in the KDRI study. Hence, grouping structures exist in donor age, peak PRA, serum creatinine, organ sharing, HLA-B and HLA-DR mismatch score, years of RRT, and primary diagnosis. Thirteen extra variables are added. The initial model therefore includes 47 variables.

To account for the center effect and the competing risk, we fit the penalized stratified and the penalized marginal models with four penalties: LASSO, ALASSO, SCAD, and MCP. Grouped variables are selected in or out together. Since the center number and the center size are both large, we consider both the regularly and the highly stratified PSH models. 
Table \ref{table 5} presents selected donor and transplant factors and corresponding estimated coefficients. Standard errors are included in the parenthesis. Refer to Appendix C for complete selection results. 
\newpage
\begin{landscape}
\begin{table}
\caption{Variable selection results and estimated regression coefficients using UNOS data. Standard errors are in the parenthesis.}
\centering
\label{table 5}
\scalebox{0.6}{
\begin{threeparttable}[b]
\begin{tabular}{lllllllllllll}
\hline\noalign{\smallskip}
 & \multicolumn{4}{c}{Regularly Stratified} & \multicolumn{4}{c}{Highly Stratified} & \multicolumn{4}{c}{Marginal} \\
& LASSO & ALASSO & SCAD & MCP & LASSO & ALASSO & SCAD & MCP & LASSO & ALASSO & SCAD & MCP \\
\noalign{\smallskip}\hline\noalign{\smallskip}
{\bf Donor Factors} &  &  &  &  &  &  &  &  &  &  &  &  \\
Age -40 yrs &  0.012 (0.001) & 0.013 (0.001) & 0.013 (0.001) & 0.012 (0.001) & 0.011 (0.001) & 0.013 (0.001) & 0.013 (0.001) & 0.012 (0.001) & 0.012 (0.001) & 0.013 (0.001) & 0.012 (0.001) & 0.012(0.001) \\
(applies to all ages)  &  &  &  &  &  &  &  &  &  &  &  &  \\
Age -18 yrs & -0.010 (0.008) & -0.013 (0.006) & -0.005 (0.007) & -0.005 (0.007) & -0.015 (0.006) & -0.014 (0.005) & -0.005 (0.005) & -0.005 (0.005) & -0.014 (0.008) & -0.018 (0.007) & -0.014 (0.008) & -0.014 (0.008) \\
(applies only if age\textless18)  &  &  &  &  &  &  &  &  &  &  &  &  \\
Age -50 yrs  & 0.020 (0.003) & 0.018 (0.003) & 0.018 (0.003) & 0.019 (0.003) & 0.018 (0.003) & 0.016 (0.002) & 0.018 (0.002) & 0.019 (0.002) & 0.020 (0.003) & 0.018 (0.003) & 0.020 (0.003) & 0.020 (0.003) \\
(applies only if age\textgreater50)  &  &  &  &  &  &  &  &  &  &  &  &  \\
Afrian American race  & 0.170 (0.031) & 0.167 (0.031) & 0.172 (0.031) & 0.168 (0.031) & 0.151 (0.023) & 0.148 (0.027) & 0.174 (0.026) & 0.169 (0.027) & 0.164 (0.033) & 0.157 (0.033) & 0.166 (0.033) & 0.166 (0.033) \\
Male & -0.037 (0.026) & - & - & - & -0.042 (0.018) & - & - & - & -0.035 (0.026) & -0.027 (0.023) & -0.034 (0.026) & -0.034 (0.026) \\
Height: per 10 cm increase  & -0.021 (0.013) &  & -0.031 (0.011) & -0.029 (0.011) & -0.008 (0.01) &  & -0.031 (0.008) & -0.028 (0.009) & -0.018 (0.014) & - & -0.019 (0.014) & -0.019 (0.014) \\
Weight & -0.227 (0.06) & -0.311 (0.052) & -0.248 (0.059) & -0.241 (0.059) & -0.169 (0.048) & -0.215 (0.043) & -0.251 (0.05) & -0.244 (0.045) & -0.213 (0.064) & -0.235 (0.059) & -0.220 (0.064) & -0.22 (0.064) \\
(per 5 kg increase if \textless 80 kg)  &  &  &  &  &  &  &  &  &  &  &  &  \\
Donation after cardiac death & - & - & - & - &- & - & - & - & - & - & - & - \\
Cause of  death: stroke & 0.061 (0.025) & 0.073 (0.024) & - & 0.067 (0.024) & 0.067 (0.021) & 0.040 (0.019) &  & 0.069 (0.022) & 0.066 (0.025) & 0.053 (0.025) & 0.066 (0.025) & 0.066 (0.025) \\
Serum creatinine-1 & 0.195 (0.036) & 0.174 (0.035) & 0.183 (0.035) & 0.186 (0.035) & 0.150 (0.026) & 0.107 (0.028) & 0.183 (0.022) & 0.187 (0.029) & 0.192 (0.035) & 0.155 (0.035) & 0.196 (0.035) & 0.196 (0.035) \\
(applies to all Cr values )  &  &  &  &  &  &  &  &  &  &  &  &  \\
Serum creatinine-1 & -0.189 (0.05) & -0.165 (0.049) & -0.176 (0.049) & -0.179 (0.049) & -0.143 (0.034) & -0.099 (0.036) & -0.177 (0.032) & -0.180 (0.036) & -0.185 (0.046) & -0.148 (0.046) & -0.189 (0.046) & -0.189 (0.046) \\
(applies if Cr\textgreater1.5 )  &  &  &  &  &  &  &  &  &  &  &  &  \\
Diabetic & 0.233 (0.047) & 0.237 (0.047) & 0.228 (0.047) & 0.232 (0.047) & 0.212 (0.039) & 0.191 (0.038) & 0.231 (0.034) & 0.235 (0.036) & 0.242 (0.05) & 0.229 (0.049) & 0.245 (0.05) & 0.245 (0.05) \\
Hypertensive & 0.137 (0.026) & 0.139 (0.026) & 0.151 (0.026) & 0.138 (0.026) & 0.132 (0.02) & 0.130 (0.021) & 0.150 (0.018) & 0.136 (0.022) & 0.142 (0.027) & 0.143 (0.027) & 0.142 (0.027) & 0.142 (0.027) \\
Cigarette users & 0.027 (0.021) & - & - & - & 0.017 (0.013) & - & - & - & 0.024 (0.018) & - & 0.025 (0.018) & 0.025 (0.018) \\
Positive HCV status & 0.098 (0.068) & - & - & - & 0.070 (0.048) & - & - & - & 0.133 (0.064) & 0.055 (0.064) & 0.135 (0.064) & 0.135 (0.064) \\
{\bf Transplant Factors}  &  &  &  &  &  &  &  &  &  &  &  &  \\
Cold ischemic time& 0.003 (0.001) & - & - & - & 0.002 (0.001) & - & - & - & 0.004 (0.002) & 0.002 (0.001) & 0.004 (0.002) & 0.004 (0.002) \\
(ref=20 hr)  &  &  &  &  &  &  &  &  &  &  &  &  \\
Organ sharing (ref=local)  &  &  &  &  &  &  &  &  &  &  &  &  \\
Regional & 0.061 (0.037) & - & - & - & 0.045 (0.022) & - & - & - & 0.030 (0.038) & - & 0.031 (0.038) & 0.031 (0.038) \\
National & -0.052 (0.031) & - & - & - & -0.038 (0.027) & - & - & - & -0.049 (0.037) & - & -0.051 (0.037) & -0.051 (0.037) \\
HLA-B mismatch  &  &  &  &  &  &  &  &  &  &  &  &  \\
\ \ 0 (ref=2 B MM) & -0.155 (0.034) & -0.176 (0.031) & -0.174 (0.031) & -0.174 (0.031) & -0.141 (0.028) & -0.134 (0.025) & -0.174 (0.027) & -0.174 (0.03) & -0.154 (0.039) & -0.158 (0.039) & -0.156 (0.039) & -0.156 (0.039) \\
\ \ 1 &  -0.068 (0.023) & -0.070 (0.023) & -0.069 (0.023) & -0.069 (0.023) & -0.057 (0.019) & -0.051 (0.021) & -0.067 (0.018) & -0.067 (0.017) & -0.055 (0.025) & -0.052 (0.024) & -0.056 (0.025) & -0.056 (0.025) \\
HLA-DR mismatch &  &  &  &  &  &  &  &  &  &  &  &  \\
\ \ 0 (ref=1 DR MM) & -0.097 (0.027) & -0.103 (0.027) & -0.102 (0.027) & -0.102 (0.027) & -0.087 (0.024) & -0.078 (0.02) & -0.102 (0.024) & -0.102 (0.023) & -0.091 (0.03) & -0.084 (0.029) & -0.092 (0.03) & -0.092 (0.03) \\
\ \ 2 & 0.055 (0.023) & 0.057 (0.023) & 0.058(0.023) & 0.057 (0.023) & 0.051 (0.018) & 0.047 (0.019) & 0.058 (0.017) & 0.057 (0.02) & 0.063 (0.025) & 0.058 (0.025) & 0.063 (0.025) & 0.063 (0.025) \\
Transplant year & -0.155 (0.021) & -0.161 (0.021) & -0.165 (0.021) & -0.163 (0.021) & -0.139 (0.018) & -0.139 (0.021) & -0.166 (0.02) & -0.164 (0.02) & -0.151 (0.025) & -0.156 (0.025) & -0.152 (0.025) & -0.151 (0.025) \\
{\it en bloc} transplant & -0.313 (0.117) & -0.285 (0.116) & -0.317 (0.117) & -0.312 (0.117) & -0.209 (0.094) & -0.083 (0.085) & -0.310 (0.094) & -0.304 (0.081) & -0.295 (0.124) & -0.132 (0.122) & -0.304 (0.124) & -0.304 (0.124) \\
Double kidney transplant & -0.334 (0.093) & -0.329 (0.092) & -0.332 (0.092) & -0.330 (0.092) & -0.247 (0.063) & -0.181 (0.047) & -0.331 (0.067) & -0.330 (0.061) & -0.329 (0.078) & -0.250 (0.078) & -0.338 (0.078) & -0.338 (0.078) \\
ABO identical & - & - & - & - & - & - & - & - & -0.023 (0.052) & - & -0.026 (0.052) & -0.027 (0.052) \\
\noalign{\smallskip}\hline
\end{tabular}
\begin{tablenotes}
\item Cr: serum creatinine.
 HLA: human leukocyte antigen.
  \end{tablenotes}
 \end{threeparttable}}
\end{table}
\end{landscape}

For the penalized stratified model, the regularly and the highly stratified select the same variables, with slight differences in estimated effects of covariates. The three oracle penalties, ALASSO, SCAD, and MCP, select similar models, in which 16 out of 18-19 factors are selected by all of them. LASSO selects 27 factors, producing larger models than the others. 
The penalized marginal model selects larger models than the penalized stratified model,  suggesting that the factors that are significant for the marginal cumulative incidence function may not necessarily have effects on the center-specific cumulative incidence function. Across penalties, LASSO, SCAD, and MCP have similar selection results and ALASSO selects a smaller model than the others.
%Take the models penalized by MCP as an example: donor's gender, cigarette usage, HCV status, cold ischemic time, organ sharing type, ABO status, recipient's gender, blood transfusion status, height, years of RRT, status of angina pectoris, and diabetes status are important factors for the marginal cumulative incidence function but not the center-specific cumulative incidence function.

 One can choose between the penalized marginal model and the penalized stratified model based on study interests. If the study interest is on the study population, then the penalized marginal model is appropriate, because patients are considered as a representative random sample from the kidney transplant population, and the effects of selected factors have a population-average interpretation \citep{conmarg}. 
Alternatively, the penalized stratified model can be applied to account for the varying patient populations across centers. In our analysis, the stratified approach was adopted as in the KDRI study.

The prognostic ability of the selected stratified PSH models in Table \ref{table 5} can be assessed by a class of discrimination and calibration methods \citep{cindex, prederror}. The concordance index (C-index) is a measure of discrimination ranging from 0 to 1 with 1 being perfect discrimination. It is defined as the proportion of all evaluable ordered patients pairs for which prediction and outcomes are concordant. \cite{cindex} adapted this measure to the competing risks setting and defined an ordered pair as evaluable if the first patient has graft failure at a time point when the second has not experienced any event or has died with a functioning graft. An evaluable ordered pair is concordant if the first patient has higher risk prediction than the second.
An alternative measure of discrimination is the D-index, originally proposed by \cite{dindex} in standard survival analysis. The D-index measures the separation of cumulative incidence curves and can be interpreted as the log-hazard ratio comparing two equal-sized prognostic groups based on dichotomizing the linear predictor from the fitted model. This index is non-negative and a larger value suggests greater discrimination ability. Prediction error (PE) of selected models can be computed as an integrated difference between the observed and the predicted cumulative incidence functions up to a time point \citep{prederror}. Using the results from \cite{stratifiedprognostic}, we extend the three measures to the stratified PSH model. 
The measure of PE is infeasible for the highly stratified PSH model, because the estimator for the baseline cumulative subdistribution hazard can not be obtained due to small strata sizes \citep{zhoustratified}. 

We randomly sample $4/5$ of the data set as a training set and the rest is a testing set. 
For regular stratification, the sampling is conducted within centers.
For high stratification, one center is regarded as one unit and the sampling is performed upon centers. We obtain parameter estimates using the training set only and then compute the C-index, the D-index, and the 5-year prediction error for the testing set. The splitting, estimation and prediction are repeated for 100 times. The averages of the three measures are reported.

Table \ref{table6} shows the prognostic statistics for each model. Standard errors are included in the parentheses. All models have similar discrimination power and prediction accuracy. ALASSO and MCP have the highest C-index. LASSO has the highest D-index and the lowest 5-year prediction error. 
We adopt the regularly stratified PSH model penalized by MCP, because it has satisfactory prognostic ability and model parsimony. Although ALASSO and MCP have similar prognostic statistics and the former has one donor factor (height) less than the latter, it is more reasonable to include both height and weight in the model, since the two are highly correlated and patients' heights can be easily obtained.

\begin{table}[t]
\caption{Prognostic statistics for penalized stratified models.  Standard errors are in the parenthesis.}
\label{table6}
\resizebox{\textwidth}{!}{
\begin{tabular}{ll>{\centering}m{1.5cm}>{\centering}m{2cm}ccc}
\hline\noalign{\smallskip}
Stratification & Penalty & No. donor factors & No. transplant factors & C-index & D-index & 5-year prediction error \\
\noalign{\smallskip}\hline\noalign{\smallskip}
Regular & LASSO & 11 & 7 & 0.635 (0.007) & 0.846 (0.033) & 0.145 (0.004) \\
 & ALASSO & 7 & 5 & 0.636 (0.006) & 0.841 (0.036) & 0.146 (0.003) \\
 & SCAD & 7 & 5 & 0.635 (0.006) & 0.832 (0.033) & 0.146 (0.003) \\
 & MCP & 8 & 5 & 0.636 (0.006) & 0.842 (0.038) & 0.146 (0.004) \\
\noalign{\smallskip}\hline\noalign{\smallskip}
High & LASSO & 11 & 7 & 0.636 (0.008) & 0.846 (0.037) & - \\
 & ALASSO & 7 & 6 & 0.636 (0.008) & 0.843 (0.038) & - \\
 & SCAD & 7 & 5 & 0.636 (0.008) & 0.841 (0.037) & - \\
 & MCP & 8 & 5 & 0.636 (0.008) & 0.841 (0.037) & -\\
\noalign{\smallskip}\hline
\end{tabular}
}
\end{table}

Our prognostic model, which we refer to as the kidney donor graft failure index (KDGFI), can be constructed in the same way as the KDRI, by using the exponential of the prognostic index (PI). Here PI is a weighted sum of selected donor factors, where the weights are the estimated regression coefficients. Transplant factors are usually not included when assessing generic donor quality \citep{rao}. Since baseline subdistribution hazards are not included, KDGFI is a relative cumulative incidence or absolute risk for graft failure compared with a reference donor whose KDGFI is 1.00. The reference donor has the following characteristics: 40-year-old, non-African American race, height 170cm, weight more than or equal to 80 kg, cause of death other than cerebrovascular accident, serum creatinine 1.0 mg/dL, non-diabetic, and non-hypertensive. For a particular deceased donor kidney, its PI is calculated by summing PI components for all applicable donor characteristics, which are listed in Table \ref{table7}. The KDGFI is then the exponential of the PI.

\begin{table}[h]
\caption{KDGFI donor characteristics and model coefficients}
\label{table7}
\resizebox{\textwidth}{!}{
\begin{tabular}{lll}
\hline\noalign{\smallskip}
Donor Characteristics & Applies to & KDGFI Prognostic Index (PI)\\
\noalign{\smallskip}\hline\noalign{\smallskip}
Age (integer years) & All donors & 0.012$\times$(age-40) \\
 & Donor with age \textless18 & -0.005$\times$(age-18) \\
 & Donor with age \textgreater50 & 0.019$\times$(age-50) \\
Height (cm) & All donors & -0.029$\times$(height-170)/10 \\
Weight (kg) & All donors with weight \textless80 kg & -0.241$\times$(weight-80)/5 \\
Ethnicity & African American donors & 0.168 \\
History of Hypertension & Hypertensive donors & 0.138 \\
History of Diabetes & Diabetic donors & 0.232 \\
Cause of Death (COD) & Donors with COD=cerebrovascular accident & 0.067 \\
Serum Creatinine & All donors & 0.186$\times$(creat-1) \\
 & Donors with creat \textgreater 1.5mg/dL & -0.179 $\times$(creat-1.5)\\
\noalign{\smallskip}\hline
\end{tabular}}
\end{table}

All donor factors in KDGFI are also included in the KDRI. These factors have the same signs but different magnitudes in the two indices. Two KDRI factors, donation after cardiac death (DCD) and HCV status, are not included the KDGFI. Although use of DCD kidneys is associated with increased risk of delayed graft function, there is no significant difference in long-term outcomes between DCD and non-DCD kidneys \citep{dcd1, dcd2}. This might be the reason that DCD is not selected. HCV status is not included in the KDGFI, because use of HCV antibody positive donor (HCVD+) kidneys does not significantly increase the risk of graft failure in HCV antibody positive recipients (HCVR+) \citep{hcv0}, and the majority of HCVD+ kidneys (77\% in the UNOS data) are transplanted to HCVR+.

%The reason HCV status is not included is that use of HCV antibody positive donors (HCVD+) kidneys significantly elevated recipient mortality\cite{hcv1, hcv2, hcv3}, but does not significantly increase the graft failure in HCV antibody positive recipients (HCVR+) \cite{hcv0}. Among the transplantations of HCVD+ kidneys,  77\% are HCVR+, composing the majority. 

\section{Discussion}
\label{sec5}
In the analysis of kidney transplant data, variable selection procedures are complicated by the presence of within-center correlations and competing risks. To take the two issues into account, this paper proposes to penalize the stratified and the marginal PSH models, where primary interests are simultaneous variable selection and parameter estimation. Within-center correlations are not explicitly modeled.
The penalized marginal model treats the whole center as one unit and achieves good performance when the number of centers is large. 
The penalized stratified model performs well when either the center size or the center number is large.
However, as seen in Table \ref{table 2} this approach may not be suitable in the analysis of matched data, for which the center size is 2. By applying the proposed methods to the UNOS data from deceased donor kidney transplants, we develop the KDGFI comprised of eight donor characteristics for assessing the relative  risk of graft failure. Our analysis results offer a new perspective in quantifying the quality of a kidney organ by considering the competing risk of DWFG,  and provide a useful tool in practical kidney transplant research.

The penalized PSH frailty model can be applied to account for the center effect when within-center correlations are of genuine interest, but it may not be appealing in our case where the primary interest is covariate effects. As previously mentioned it is restrictive regarding the type of dependency encompassed and requires the center effect to act proportionally on the subdistribution hazard.  Additionally, calculation is cumbersome when the center number is large, due to the involvement of double iterative procedures \citep{PrenticeMultivariate}. The first iteration calculates frailties and the number of parameters to be estimated is the number of centers. The second iteration conducts the penalization procedure conditional on the frailties. In contrast, the stratified approach only involves the penalization procedure.
In the case of the KDGFI study, the number of transplant centers is 256. Using the frailty model requires extra estimation of 256 parameters, which considerably increases computing time.

%At time $t$, the counting process counts all individuals who have not failed from any events or have been dead with a functioning graft over the whole population. At time $t$, the counting is conducted within a center, which considerably shortens computing time. 

The current study was conducted under the setting where $d < K$ or/and $d < n_k$  for the stratified approach and $d < K$ for the marginal approach, while in many applications, data may be high-dimensional, for example, $d > n$, $n > d > n_k > K$, or $n > d > K > n_k$. Due to the curse of dimensionality, computation may be prohibitive \citep{overview}. The proposed penalization methods may not accurately select the underlying model, and their theoretical properties may not be valid. Future work could involve extending the proposed methods to the high-dimensional setting, along with developing efficient and robust computing procedures.

\section*{Acknowledgments}
This publication was made possible by CTSA Grant Number UL1 TR000142 from the National Center for Advancing Translational Science (NCATS), a component of the National Institutes of Health (NIH).  Its contents are solely the responsibility of the authors and do not necessarily represent the official view of NIH.
This work was supported in part by Health Resources and Services Administration contract 234-2005-37011C. The content is the responsibility of the authors alone and does not necessarily reflect the views or policies of the Department of Health and Human Services, nor does mention of trade names, commercial products, or organizations imply endorsement by the U.S. Government.

\bibliographystyle{plainnat}

%\bibliography{ref}

\newpage
\begin{appendices}

\section{Large Sample Properties of the Penalized Stratified PSH Model}
\label{thm1}
For simplicity, we shall work on the finite time interval $[0, \tau]$. Because of the distinct asymptotic behaviors of the center number and the center size, we write $\tbeta^S$ as $\tbeta_r^S$ and $\tbeta_h^S$ for regularly and highly stratified data respectively. The following conditions are needed to establish the oracle properties.
\begin{enumerate}[]
\item[A1.] $\int_0^{\tau}\lambda_{k0}(t)dt < \infty$ for $k=1, \dots, K$.
\item[A2.] For regularly stratified data, $\{N_{ki}(.), Y_{ki}(.), \bz_{ki}(.)\}_{i=1, \dots, n_k}$ are independently and identically distributed; For highly stratified data, $\{N_{ki}(.), Y_{ki}(.), \bz_{ki}(.), i=1, \dots, n_k\}_{k=1, \dots, K}$are independently distributed.
\item[A3.] Define 
\begin{equation}
s_k^{(p)}(\bbeta, t) = \lim_{n_k \to \infty}n_k^{-1}\sum_{i=1}^{n_k}I(C_{ki}\geq t)Y_{ki}(t)\bz_{ki}^{\otimes p}\exp(\bbeta^T\bz_{ki}) 
\end{equation}
\begin{equation}
S_k^{(p)}(\bbeta, t) = n_k^{-1}\sum_{i=1}^{n_k}I(C_{ki}\geq t)Y_{ki}(t)\bz_{ki}^{\otimes p}\exp(\bbeta^T\bz_{ki})
\end{equation}
\begin{equation}
\hS_k^{(p)}(\bbeta, t) = n_k^{-1}\sum_{i=1}^{n_k}\hw_{ki}(t)Y_{ki}(t)\bz_{ki}^{\otimes p}\exp(\bbeta^T\bz_{ki})
\end{equation}
For regularly stratified data, there exits a neighborhood $\B$ of $\bbeta_0$ such that the following conditions are satisfied: (i) there exits a scalar, a vector, and a matrix function $\szero_k, \sone_k$, and $\stwo_k$ defined on $\B \times [0, \tau]$ such that $\text{sup}_{t\in [0, \tau], \bbeta \in \B}\|\Sp_k(\bbeta, t)-\spp_k(\bbeta, t)\| \to 0$ in probability, $p=0, 1, 2$; (ii). the matrix $\bOmega_k=\int_0^{\tau}\vt_k(\bbeta_0, t)\szero_k(\bbeta_0, t)\lambda_{0k}(t)dt$ is positive definite, where $\vt_k=\stwo_k/\szero_k-(\sone_k/\szero_k)(\sone_k/\szero_k)^T$.

For highly stratified data, the matrix $\bI(\bbeta_0)=\lim_{K\to \infty}K^{-1}\sum_{k=1}^K\E[\sum_{i=1}^{n_k}\int_0^{\tau}\{\bz_{ki}-\overline \bz_{k}(\bbeta_0, u)\}^{\otimes 2}\Szero(\bbeta_0, u)\lambda_{0k}(u)du$ is positive definite.

\item[A4.] The penalty function satisfies that $a_n=O_p(n^{-1/2})$ and $b_n \to 0$.

\item[A5.] For regular stratified data, $p_{\lambda_n}(|\beta|)$ satisfies  $\lim\limits_{n_k\to \infty}\sqrt{n_k} \inf\limits_{|\beta|\leq Cn^{-1/2}}p'_{\lambda_n}(|\beta|) \to \infty, k=1, \dots, K$.

For highly stratified data, $p_{\lambda_n}(|\beta|)$ satisfies  $\lim\limits_{K\to \infty}\overline mK^{1/2} \inf\limits_{|\beta|\leq Cn^{-1/2}}p'_{\lambda_n}(|\beta|) \to \infty$, where $\overline m=\lim\limits_{K \to \infty}n/K$.
\end{enumerate}

%%%%%%%%%%%%%%%%%%%%%%%%%%%%%%%%%%%%%%%%%%%%%%
\subsection{The Regular Stratified PSH Model}
Theorem 1 states selection consistency and the oracle properties for the penalization methods for the regularly stratified PSH model.
\begin{thm}
\label{thm sr}
For regularly stratified data with finite $K$, when $n \to \infty$, $n_k \to \infty$. The following holds under Condition A.1 - A.5,
\begin{enumerate}[(a.)]
\item $\tbeta_r^S$ is a root-$n$ consistent estimator for $\bbeta_0$, i.e. $||\tbeta_r^S - \bbeta_0||=O_p(n^{-1/2})$.
\item (Oracle properties) With probability tending to 1, the root-$n$ consistent estimator $\tbeta_r^S$ satisfies
\begin{enumerate}[(i.)]
\item (Sparsity) $\tbeta^S_{2r}=\bzero$
\item (Asymptotic normality) $n^{1/2}\sum_{k=1}^K\pi_k(\bOmega_{11k}+\bP)\{\tbeta^S_{1r} -\bbeta_{01}+(\bOmega_{11k}+\bP)^{-1}\bb\} \to N(\bzero, \sum_{k=1}^K\pi_k\bSigma_{11k})$, where $p_k=n_k/n \to \pi_k$, $\bOmega_{11k}$ and $\bSigma_{11k}$ are the first $s \times s$ submatrix of $\bOmega_k(\bbeta_{0})$ and $\bSigma_k(\bbeta_{0})$, defined in the Appendix.
\end{enumerate}
\end{enumerate}
\end{thm}
%%%%%%%%%%%%%%%%%%%%%%%%%%%%%%%%%%%%%%%%
\subsubsection{Proof of Theorem 1.a}
\label{them1a}
To prove $\tbeta_r^S - \bbeta_0=O_p(n^{-1/2})$, it is sufficient to prove that for any positive $\epsilon$, there exists a large constant $C$ such that 
\begin{align}
\label{0}
Pr\{\sup_{\|\bu\|=C} Q^S(\bbeta_0+\alpha_n\bu) < Q^S(\bbeta_0)\}\geq 1-\epsilon
\end{align}
where $\alpha_n=n^{-1/2}+a_n$. 
This means with probability at least $1-\epsilon$, there exits a local maximum such that $\|\tbeta_r^S-\bbeta_0\|=O_p(\alpha_n)$. 

Denoting $D_n(\bu)=\dfrac{1}{n}\{Q^S(\bbeta_0+\alpha_n\bu)-Q^S(\bbeta_0)
\}$, we have
\begin{align}
\label{1}
D_n(\bu) &\leq \dfrac{1}{n}\sum_{k=1}^K\left [\{l_k(\bbeta_0+\alpha_n\bu)-l_k(\bbeta_0)\}-n_k\sum_{j=1}^s \{p_{\lambda_n}(|\beta_{j0}+\alpha_nu_j|)-p_{\lambda_n}(|\beta_{j0}|)\}\right ]\\
&=\dfrac{1}{n}\sum_{k=1}^Kn_kD_{n_k}(\bu)
\end{align}
where $D_{n_k}(\bu)=\dfrac{1}{n_k}\{l_k(\bbeta_0+\alpha_n\bu)-l_k(\bbeta_0)\}-\sum_{j=1}^s \{p_{\lambda_n}(|\beta_{j0}+\alpha_nu_j|)-p_{\lambda_n}(|\beta_{j0}|)\} := D_1+D_2$

By Taylor expansion, $D_1$ is 
\begin{align}
\dfrac{1}{n_k}\{l_k(\bbeta_0+\alpha_n\bu)-l_k(\bbeta_0)\} &= \dfrac{1}{n_k}(\dfrac{\partial l_k(\bbeta_0)}{\partial \bbeta})^T\alpha_n\bu - \dfrac{1}{2}\bu^T\{-\dfrac{1}{n_k}\dfrac{\partial^2 l_k(\bbeta_0)}{\partial \bbeta^T\partial \bbeta}+o_p(1)\}\bu\alpha_n^2\\
&= O_p(n_k^{-1/2})\alpha_n\bu-\dfrac{1}{2}\alpha_n^2\bu^T\{\bOmega_k(\bbeta_0)+o_p(1)\}\bu\\
&=O_p(Cn_k^{-1/2}\alpha_n)+O_p(C^2\alpha_n^2)
\end{align}
From \citet{scadcox}, $D_2$ is bounded by $\sqrt{s}a_n\alpha_n||\bu||+b_n\alpha_n^2||\bu^2||$, which has the order of $C\alpha_n^2$ if $b_n\to 0$. Define $r_k=n_k/n \to \pi_k$, which is a finite number between $0$ and $1$. Summing over $K$ strata, 
\begin{align}
D_n(\bu) &\leq \sum_{k=1}^K\dfrac{n_k}{n}\{O_p(Cn_k^{-1/2}\alpha_n)+O_p(C^2\alpha_n^2)+O_p(C\alpha_n^2)\}\\
&=\sum_{k=1}^Kr_k\{O_p(C\alpha^2_nr_k^{-1/2})+O_p(C^2\alpha_n^2)+O_p(C\alpha_n^2)\}
\end{align}
By choosing a sufficiently large $C$, $O_p(C^2\alpha_n^2)$ dominates the others. Thus inequality(\ref{0}) holds.

%%%%%%%%%%%%%%%%%%%%%%%%%%%%%%%%%%%%%%%%%%%%%%
\subsubsection{Proof of Theorem 1.b. (i)}
\label{them1b1}
It is sufficient to show that for any given $\bbeta_{1}$ that is root-n consistent and for any constant $C$,
\begin{align}
\label{eq9}
Q^S(\bbeta_{1}, \bzero)=\max\limits_{||\bbeta_{2}||\leq Cn^{-1/2}}Q^S(\bbeta_{1}, \bbeta_{2})
\end{align}
This is equivalent to show that $\dfrac{\partial Q^S(\bbeta)}{\partial \beta_j} <0$ for $0<\beta_j<Cn^{-1/2}$, and $\dfrac{\partial Q^S(\bbeta)}{\partial \beta_j} >0$ for $-Cn^{-1/2}<\beta_j<0$ with probability tending to 1 as $n\to\infty$, $j=s+1, \dots, d$.

For stratum $k$, define the following
\begin{align}
X_k(\bbeta, \tau)&=\dfrac{1}{n_k}\{l_k(\bbeta)-l_k(\bbeta_0)\}\\
&=\dfrac{1}{n_k}\sum_{i=1}^{n_k}\int_0^{\tau}\{(\bbeta-\bbeta_0)^T\bZ_{ki}-\log\dfrac{\hSzero_k(\bbeta, u)}{\hSzero_k(\bbeta_0, u)}\}\hw_{ki}(u)dN_{ki}(u)\\
A_k(\bbeta, \tau)&=\dfrac{1}{n_k}\sum_{i=1}^{n_k}\int_0^{\tau}\{(\bbeta-\bbeta_0)^T\bZ_{ki}-\log\dfrac{\hSzero_k(\bbeta, u)}{\hSzero_k(\bbeta_0, u)}\}\hw_{ki}(u)\lambda_{1k}(u)du\\
&=\int_{0}^{\tau}\{(\bbeta-\bbeta_0)^T\hSone_k(\beta_0, u)-\hSzero_k(\bbeta_0, u)\log\dfrac{\hSzero_k(\bbeta, u)}{\hSzero_k(\bbeta_0, u)}\}\lambda_{1k0}(u)du
\end{align}
From Fine and Gray (1999), $X_k(\bbeta, .)-A_k(\bbeta, .)$ is a sum of martingale integrals with respect to locally bounded process. Since 
$\lim_{n_k \to \infty}\hSp(\bbeta, t)=G(t)\spp_k(\bbeta, t)$, 
by condition 3 it follows that for each $\beta \in \B$, 
\begin{align}
A_k(\beta, \tau) \to f_k(\bbeta, \tau)=\int_{0}^{\tau}\{(\bbeta-\bbeta_0)^T\sone_k(\beta_0, u)-\szero_k(\bbeta_0, u)\log\dfrac{\szero_k(\bbeta, u)}{\szero_k(\bbeta_0, u)}\}\lambda_{1k0}(u)du
\end{align}
where $f_k$ has the following properties
\begin{align}
f_k(\bbeta_0)=0, \ \ \dfrac{\partial f_k(\bbeta_0)}{\partial \bbeta}=0, \ \ -\dfrac{\partial^2 f_k(\bbeta_0)}{\partial \bbeta\partial \bbeta^T}=\bOmega_k(\bbeta_0)
\end{align}
Then the following holds 
\begin{align}
\dfrac{1}{n_k}\{l_k(\bbeta)-l_k(\bbeta_0)\}&=f_k(\bbeta, \tau) +O_p(\dfrac{||\bbeta-\bbeta_0||}{\sqrt{n_k}})
\end{align}
We also have $f_k(\bbeta)=-\dfrac{1}{2}(\bbeta-\bbeta_0)^T\{\Omega_k(\bbeta_0)+o(1)\}(\bbeta-\bbeta_0)$.
By Taylor expansion, we obtain
\begin{align}
\dfrac{\partial Q^S(\bbeta)}{\partial\beta_j}&=\sum_{k=1}^K\dfrac{\partial l_k(\bbeta)}{\partial \beta_j}-n_kp'_{\lambda_n}(|\beta_j|)sgn(\beta_j)\\
&=\sum_{k=1}^Kn_k\sum^d_{l=1}\dfrac{\partial^2 f_k(\bbeta_0)}{\partial\beta_j \partial \beta_l}(\beta_l - \beta_{l0}) +O_p(n_k||\bbeta-\bbeta_0||^2)-n_kp'_{\lambda_n}(|\beta_j|)sgn(\beta_j)\\
&=\sum_{k=1}^KO_p(n_kn^{-1/2})+O_p(r_k)-n_kp'_{\lambda_n}(|\beta_j|)sgn(\beta_j)\\
&=\sum_{k}^Kn_k^{1/2}\{O_p(r_k)-n_k^{1/2}p'_{\lambda_n}(|\beta_j|)sgn(\beta_j)\}
\end{align}
Since $\lim\limits_{n_k\to \infty} \sqrt{n_k} \inf\limits_{||\beta||\leq Cn^{-1/2}}p'_{\lambda_n}(|\beta|) \to \infty, k=1, \dots, K$ in condition C.2, the sign of $\dfrac{\partial Q^S(\bbeta)}{\beta_j}$ is completely determined by the sign of $\beta_j$. 
%%%%%%%%%%%%%%%%%%%%%%%%%%%%%%%%%%%%%%%%
\subsubsection{Proof of Theorem 1.b. (ii)}
\label{them1b2}
It is easy to show that there exists a root-n consistent local maximizer $\tbeta_{r1}^S$ of $Q(\bbeta_{1}, \bzero)$, satisfying equation
$\dfrac{\partial Q(\bbeta)}{\partial\bbeta_{1}}\Big|_{\bbeta=((\tbeta_{r1}^S)^T, \bzero)^T}=0$.
Denote $U^S(\bbeta)$ and $H^S(\bbeta)$ as the score function and the Hessian matrix of $l^S(\bbeta)$ and they are
\begin{align}
U^S(\bbeta)&=\sum_{k=1}^KU_{k}(\bbeta)=\sum_{k=1}^K\sum_{i=1}^{n_k}\int_0^{\infty}\{\bZ_{ik}-\dfrac{\hSone_i(\bbeta, u)}{\hSzero_i(\bbeta, u)}\}\hw_{ik}(u)dN_{ik}(u)\\
H^S(\bbeta)&=\sum_{k=1}^KH_{k}(\bbeta)=\sum_{k=1}^K\sum_{i=1}^{n_k}\int_0^{\infty}\left [\dfrac{\hStwo_i(\bbeta, u)}{\hSzero_i(\bbeta, u)}-\{\dfrac{\hSone_i(\bbeta, u)}{\hSzero_i(\bbeta, u)}\}^{\otimes2}\right ]\hw_{ik}(u)dN_{ik}(u)
\end{align}
For $j=1, \dots, s$
\begin{equation}
\begin{split}
0&=\sum_{k=1}^K\dfrac{\partial l_k(\bbeta)}{\partial\beta_j}\Big|_{\bbeta=(\tbeta_{1}^T, \bzero)^T}-n_kp'_{\lambda_n}(|\tilde\beta^S_{rj}|)sgn(\tilde\beta^S_{rj})\\
&=\sum_{k=1}^K\dfrac{\partial l_k(\bbeta_0)}{\partial \beta_j}-\sum_{l=1}^s\{-\dfrac{\partial^2 f_k(\bbeta_0)}{\partial\beta_j \partial \beta_l}+o_p(1)\}(\tilde\beta^S_{rl}-\beta_{l0})-\\
&\quad  n_k\left(p'_{\lambda_n}(|\beta_{j0}|)sgn(\beta_{j0})+\{p''_{\lambda_n}(| \beta_{j0}|)+o_p(1)\}(\tilde\beta^S_{rj}-\beta_{j0})\right)
\end{split}
\end{equation}
Denote $U_{11k}(\bbeta)$ as the first $s$ elements of the score function $U_k(\bbeta)$ for stratum $k$. As $n_k \to \infty$, $n_k^{-1/2}U_{11k}(\bbeta_0) \to N(\bzero, \bSigma_{11k}(\bbeta_0))$ in distribution, where $\bSigma_{11k}$ is the first $s\times s$ submatrix of $\bSigma_k$. Th explicit formula of $\bSigma_k$ can be found in Fine and Gray(1999). Thus $n^{-1/2}\sum_{k=1}^KU_{11k}(\bbeta_0)= n^{-1/2}\sum_{k=1}^Kr_k^{1/2}n_k^{-1/2}U_{11k}(\bbeta_0) \to N(\bzero, \Sigma_{11}(\bbeta_0))$, where $\Sigma_{11}=\sum_{k=1}^K\pi_k\Sigma_{11k}$. Furthermore, let $H_{11k}$ be the first $s\times s$ submatrix of of $H_k$ and $-\dfrac{1}{n_k}H_{11k} \to \bOmega_{11k}(\bbeta_0)$ in probability, where $\bOmega_{11k}$ is the first $s\times s$ submatrix of $\bOmega_k$. Let $\bb_1$ be the first $s$ elements of $\bb$, and $\bP_{11}$ be the first $s \times s$ submatrix of $\bP$. By Slutsky's Theorem, 
\begin{align}
n^{1/2}\sum_{k=1}^K\pi_k(\bOmega_{11k}+\bP_{11})\{\tbeta^S_{r1} -\bbeta_{01}+(\bOmega_{11k}+\bP_{11k})^{-1}\bb_1\} \to N(\bzero, \bSigma_{11}(\bbeta_0))
\end{align}
This completes the proof. 

%%%%%%%%%%%%%%%%%%%%%%%%%%%%%%%%%%%%%%%%%%%%%%
\subsection {The highly stratified PSH Model}
Theorem 2 states selection consistency and the oracle properties of the penalization methods for the highly stratified PSH model.
\label{thm2}
\begin{thm}
\label{thm sh}
For highly stratified data with finite $n_k$, when $n \to \infty$, $K \to \infty$ . The following holds under Condition A.1 - A.5,
\begin{enumerate}[(a.)]
\item $\tbeta_h^S$ is a root-$n$ consistent estimator for $\bbeta_0$, i.e. $||\tbeta_h^S - \bbeta_0||=O_p(n^{-1/2})$.
\item (Oracle properties) With probability tending to 1, the root-$n$ consistent estimator $\tbeta_h^S$ satisfies
\begin{enumerate}[(i.)]
\item(Sparsity) $\tbeta^S_{2h}=\bzero$
\item(Asymptotic normality) $n^{1/2}(\bOmega_{11}+\bP)\{\tbeta^S_{1h} -\bbeta_{01}+(\bOmega_{11}+\bP)^{-1}\bb\} \to N(\bzero, \overline m^{-1}\bSigma_{11}(\bbeta_0))$, where $\overline m=\lim\limits_{K\to \infty}n/K$, $\bOmega_{11}$ and $\bSigma_{11}$ are the first $s \times s$ submatrix of $\bOmega(\bbeta_{0})$ and $\bSigma(\bbeta_{0})$, defined in the Appendix.
\end{enumerate}
\end{enumerate}
\end{thm}

Following the proof of Theorem 1, we briefly give the proof of Theorem 2.
%%%%%%%%%%%%%%%%%%%%%%%%%%%%%%%%%%%%%%%%%%%%
\subsubsection{Proof of Theorem 2.a. }
When data are highly-stratified, the strata size $n_k$ is finite and $K\to \infty$ as $n \to \infty$. To prove the validity of equation (\ref{0}), we also write $D_n(\bu)=\dfrac{1}{n}\{Q^S(\bbeta_0+\alpha_n\bu)-Q^S(\bbeta_0)\}$, and 
\begin{align}
\label{high1}
D_n(\bu) &\leq \dfrac{1}{n}\sum_{k=1}^K\{l_k(\bbeta_0+\alpha_n\bu)-l_k(\bbeta_0)\}-\sum_{j=1}^s \{p_{\lambda_n}(|\beta_{j0}+\alpha_nu_j|)-p_{\lambda_n}(|\beta_{j0}|)\}\\
&:=I_1+I_2
\end{align}
By Taylor expansion, 
\begin{align}
I_1&=\dfrac{1}{n}\sum_{k=1}^K(\dfrac{\partial l_k(\bbeta_0)}{\partial \bbeta})^T\alpha_n\bu - \dfrac{1}{2}\bu^T\{-\dfrac{1}{n}\sum_{k=1}^K\dfrac{\partial^2 l_k(\bbeta_0)}{\partial \bbeta^T\partial \bbeta}+o_p(1)\}\bu\alpha_n^2\\
&=O_p(\overline m^{-1/2}\alpha_nC)+O_p(\alpha_n^2C^2)
\end{align}
where $\overline m=n/K$ is the average strata size as $n \to \infty$ and is a finite number. 
From Section \ref{them1a}, term $I_2$ has the order of $C\alpha_n^2$ if $b_n\to 0$. 
By choosing a sufficiently large $C$, $O_p(C^2\alpha_n^2)$ dominates the others. Thus inequality(\ref{0}) holds.

%%%%%%%%%%%%%%%%%%%%%%%%%%%%%%%%%%%%%%%%%%%%
\subsubsection{Proof of Theorem 2.b. (i)}
Following Section \ref{them1b1}, we prove equation (\ref{eq9}) also holds when data are highly stratified. Based on the results from Zhou et al. (2001), the following holds 
\begin{align}
\dfrac{1}{K}\{l^S(\bbeta)-l^S(\bbeta_0)\}&=\dfrac{1}{K}\sum_{k=1}^{K}\left [\sum_{i=1}^{n_k}\int_0^{\infty}(\bbeta-\bbeta_0)^T\bZ_{ki}dN_{ki}(u)-\int_0^{\infty}\log\dfrac{S_k^{(0)}(\bbeta, u)}{S_k^{(0)}(\bbeta_0, u)}d\overline N_k(u)\right ]+o_p(1)\\
&=\dfrac{1}{K}\sum_{k=1}^{K}X_k(\bbeta)+o_p(1)
\end{align}
where $S_k^{(p)}(\bbeta, u)=\dfrac{1}{n_k}\sum_{i=1}^{n_k} w_{ik}(u)Y_{ik}(u)\bZ_{ik}(u)^{\otimes p} \exp\{\bbeta^T\bZ_{ik}\}$ and $d\overline N_k(u)=\sum_{i=1}^{n_k}dN_{ki}(u)$. Correspondingly, 
\begin{align}
X_k(\bbeta)&=\sum_{i=1}^{n_k}\int_0^{\infty}(\bbeta-\bbeta_0)^T\bZ_{ki}dN_{ki}(u)-\int_0^{\infty}\log\dfrac{S_k^{(0)}(\bbeta, u)}{S_k^{(0)}(\bbeta_0, u)}d\overline N_k(u),\\
\chi_k(\bbeta)&=\E\{X_k(\bbeta)\}, \text{and} \ \ \chi(\bbeta)=\lim \limits_{K\to \infty}\dfrac{1}{K}\sum_{k=1}^{K}\chi_k(\bbeta)\\
\chi(\bbeta_0)&=0, \ \ \dfrac{\partial \chi(\bbeta_0)}{\partial \bbeta}=0, \ \ -\dfrac{\partial^2 \chi(\bbeta_0)}{\partial \bbeta\partial \bbeta^T}=\bI(\bbeta_0)
\end{align}
Because $\dfrac{1}{K}\sum_{k=1}^K\E\{|X_k(\bbeta)-\chi_k(\bbeta)|^2\}=O_p(1)$, we obtain $\dfrac{1}{K}\{l^S(\bbeta)-l^S(\bbeta_0)\}=\chi(\bbeta)+O_p(K^{-1/2})$. Also We also have $\chi(\bbeta)=-\dfrac{1}{2}(\bbeta-\bbeta_0)^T\{\bI(\bbeta_0)+o(1)\}(\bbeta-\bbeta_0)$.
By Taylor expansion, we obtain
\begin{align}
\dfrac{\partial Q^S(\bbeta)}{\partial\beta_j}&=\dfrac{\partial l^S(\bbeta)}{\partial \beta_j}-np'_{\lambda_n}(|\beta_j|)sgn(\beta_j)\\
&=K\sum^d_{l=1}\dfrac{\partial^2 \chi(\bbeta_0)}{\partial\beta_j \partial \beta_l}(\beta_l - \beta_{l0}) +O_p(K^{1/2})-np'_{\lambda_n}(|\beta_j|)sgn(\beta_j)\\
&=O_p(Kn^{-1/2}+K^{1/2})-np'_{\lambda_n}(|\beta_j|)sgn(\beta_j)\\
&=K^{1/2}\{O_p(\overline m^{-1/2})-\overline mK^{1/2}p'_{\lambda_n}(|\beta_j|)sgn(\beta_j)\}
\end{align}
As $\lim\limits_{K\to \infty} \overline mK^{1/2} \inf\limits_{||\beta||\leq Cn^{-1/2}}p'_{\lambda_n}(|\beta|) \to \infty$ in condition A5, the sign of $\dfrac{\partial Q^S(\bbeta)}{\beta_j}$ is completely determined by the sign of $\beta_j$. 
%%%%%%%%%%%%%%%%%%%%%%%%%%%%%%%%%%%%%%%%%%%%
\subsubsection{Proof of Theorem 2.b. (ii)}
\label{them2b2}
There exists $\tbeta^S_{h1}$ satisfying equation
$\dfrac{\partial Q(\bbeta)}{\partial\bbeta_{1}}\Big|_{\bbeta=((\tbeta^S_{h1})^T, \bzero)^T}=0$.
For $j=1, \dots, s$
\begin{equation}
\begin{split}
0&=\dfrac{\partial l^S(\bbeta)}{\partial\beta_j}\Big|_{\bbeta=((\tbeta^S_{h1})^T, \bzero)^T}-np'_{\lambda_n}(|\tilde\beta_j|)sgn(\tilde\beta^S_{hj})\\
&=\dfrac{\partial l^S(\bbeta_0)}{\partial \beta_j}-\sum_{l=1}^s\{-\dfrac{\partial^2 l^S(\bbeta_0)}{\partial\beta_j \partial \beta_l}\}(\tilde\beta^S_{hl}-\beta_{l0})- n\left(p'_{\lambda_n}(|\beta_{j0}|)sgn(\beta_{j0})+\{p''_{\lambda_n}(| \beta_{j0}|)+o_p(1)\}(\tilde\beta^S_{hj}-\beta_{j0})\right)
\end{split}
\end{equation}
Denote $U^S_h(\bbeta)$ and $H^S_h(\bbeta)$ as the score function and the Hessian matrix for highly stratified data. The sub-vector $U^S_{h1}(\bbeta)$ is the first element of $U^S_h(\bbeta)$, and $H^S_{h11}(\bbeta)$ is the first  $s \times s$ matrix of $H^S_h(\bbeta)$. 
 As $K \to \infty$, $K^{-1/2}U^S_{h1}(\bbeta_0) \to N(\bzero, \bSigma_{11h}(\bbeta_0))$ in distribution, where $\bSigma_{11h}$ is the first $s\times s$ submatrix of $\bSigma_h$. Details of the matrix $\bSigma_h$ can be found in Zhou et al.(2011). Also $-\dfrac{1}{n}H_{h11} \to \bOmega_{h11}(\bbeta_0)$ in probability, and $\bOmega_{h11}$ is the first $s\times s$ submatrix of $\bOmega_h$. Here $ \overline m\bOmega_h$ equals to $\bI$ defined in condition A3.
By Slutsky's Theorem, 
\begin{align}
n^{1/2}(\bOmega_{11}+\bP_{11})\{\tbeta_{1} -\bbeta_{01}+(\bOmega_{11}+\bP_{11})^{-1}\bb_1\} \to N(\bzero, \overline m^{-1}\bSigma_{11}(\bbeta_0))
\end{align}
This completes the proof. 

\subsection{Oracle Properties of Penalties}
\label{oracle}
Conditions A.4 and A.5 provide guidance to select penalties that will have the oracle properties in the stratified PSH model.  With a proper choice of $\lambda$, we show that ALASSO, SCAD, and MCP satisfy the two conditions.
\begin{enumerate}[(a).]
\item  ALASSO: since \citet{zhoustratified} have shown that $\hat\bbeta$ is root-$n$ consistent, the data-adaptive weight $\theta_j=|\hbeta^S_j|^{-1}$ has the order of $n^{1/2}$. When $\sqrt{n}\lambda_n \to 0$, $\sqrt{n}a_n=\sqrt{n}\lambda_n\theta \to 0$ and $b_n=0$. Hence Condition A.4 is satisfied. For regularly stratified data, when $n_k\lambda_n \to \infty$, $\lim\limits_{n_k\to \infty}\sqrt{n_k} \inf\limits_{||\beta||\leq Cn^{-1/2}}p'_{\lambda_n}(|\beta|)=\sqrt{n_k}\lambda_n\theta =n_k\pi_k^{1/2}\lambda_n\to \infty$. For highly stratified data, when $K\lambda_n \to \infty$,  $\lim\limits_{K\to \infty}\overline mK^{1/2} \inf\limits_{||\beta||\leq Cn^{-1/2}}p'_{\lambda_n}(|\beta|)={\overline m}^{3/2}K\lambda_n \to \infty$. Condition A.5 is satisfied. 

\item  SCAD: when $\lambda_n \to 0$, $a_n=b_n=0$. Condition A.4 is satisfied. For regular stratified data,
When $\sqrt{n_k}\lambda_n \to \infty$, $\lim\limits_{n_k\to \infty}\sqrt{n_k}\inf\limits_{||\beta||\leq Cn^{-1/2}}p'_{\lambda_n}(|\beta|)=\sqrt{n_k}\lambda_n \to \infty$; for highly-stratified data, when $K^{1/2}\lambda_n \to \infty$, $\lim\limits_{K\to \infty}\overline mK^{1/2} \inf\limits_{||\beta||\leq Cn^{-1/2}}p'_{\lambda_n}(|\beta|)=$

$\overline m K^{1/2}\lambda_n \to \infty$. Condition A.5 holds.  

\item MCP: when $\lambda_n \to 0$, $a_n=b_n=0$ for sufficiently large $n$. Condition A.4 is satisfied. For regularly stratified data, when $\sqrt{n_k}\lambda_n \to \infty$, $\sqrt{n_k}\inf\limits_{||\beta||\leq Cn^{-1/2}}p'_{\lambda_n}(|\beta|)=\sqrt{n_k}\lambda_n-C\sqrt{\pi_k}/\gamma \to \infty$. 
For highly stratified data, when $K^{1/2}\lambda_n \to \infty$, 

$\overline mK^{1/2} \inf\limits_{||\beta||\leq Cn^{-1/2}}p'_{\lambda_n}(|\beta|)=\overline m K^{1/2}\lambda_n-C\overline m^{1/2}/\gamma \to \infty$. Condition A.5 holds.
\end{enumerate}
Hence, according to Theorem \ref{thm sr} and Theorem \ref{thm sh}, when $\lambda_n$ is  chosen appropriately, the three penalized estimators possess the oracle properties under both stratification regimes and asymptotically follow a normal distribution $\sqrt{n}(\tbeta^S_{1r}-\bbeta_{01}) \to N(\bzero, $
$(\sum_{k=1}^{K}\pi_k\bOmega_{11k})^{-1}(\sum_{k=1}^K\pi_k\bSigma_{11k})(\sum_{k=1}^{K}\pi_k\bOmega_{11k})^{-1})$ when data are regularly stratified, and follow $\sqrt{n}(\tbeta^S_{1h}-\bbeta_{01}) \to N(\bzero, \bOmega_{11}^{-1} \overline m^{-1}\bSigma_{11}\bOmega_{11}^{-1})$ when data are highly stratified. 
%They perform as well as $\hbeta$ for estimating $\bbeta_1$ knowing $\bbeta_2=\bzero$ in advance. 

For LASSO, the oracle properties do not hold. From Condition A.4, $a_n=\lambda_n=O_p(n^{-1/2})$ and $b_n=0$, whereas Condition A.5 requires that $n_k^{1/2}\lambda_n= \pi_k^{1/2}n^{1/2}\lambda \to \infty$ for regularly stratified data and ${\overline m}^{1/2}n^{1/2}\lambda_n \to \infty$ for highly stratified data. The two conditions can not be simultaneously held.
\end{appendices}
%%%%%%%%%%%%%%%%%%%%%%%%%%%%%%%%%%%%%%%%%%%%
\begin{appendices}
\section{Large Sample Properties of the Penalized Marginal PSH Model}
\label{thm3}
We assume the following regularity conditions 
\begin{enumerate}[B1.]
\item $\int_0^{\tau}\lambda_{10}(t)dt < \infty$
\item Define 
\begin{equation}
\spp(\bbeta, t) = \lim_{K \to \infty}K^{-1}\sum_{k=1}^K\sum_{i=1}^{n_k}I(C_{ki}\geq t)Y_{ki}(t)Z_{ki}^{\otimes p}\exp(\bbeta^T\bZ_{ki}) 
\end{equation}
\begin{equation}
\Sp(\bbeta, t) = K^{-1}\sum_{k=1}^K\sum_{i=1}^{n_k}I(C_{ki}\geq t)Y_{ki}(t)Z_{ki}^{\otimes p}\exp(\bbeta^T\bZ_{ki})
\end{equation}
\begin{equation}
\hSp(\bbeta, t) = K^{-1}\sum_{k=1}^K\sum_{i=1}^{n_k}\hw_{ki}(t)Y_{ki}(t)Z_{ki}^{\otimes p}\exp(\bbeta^T\bZ_{ki})
\end{equation}
There exits a neighborhood $\B$ of $\bbeta_0$ such that the following conditions are satisfied: 
(i) there exits a scalar, a vector, and a matrix function $\szero, \sone$, and $\stwo$ defined on $\B \times [0, \tau]$ such that $\text{sup}_{t\in [0, \tau], \bbeta \in \B}\|\Sp(\bbeta, t)-\spp(\bbeta, t)\| \to 0$ in probability, $p=0, 1, 2$; 
(ii). the matrix $\bOmega^M=\int_0^{\tau}\vt(\bbeta_0, t)\szero(\bbeta_0, t)\lambda_{10}(t)dt$ is positive definite, where $\vt=\stwo/\szero-(\sone/\szero)(\sone/\szero)^T$.
\item $\spp(\bbeta, t)$ are continuous functions of $\bbeta \in \B$ uniformly in $t \in [0. \tau]$ and are bounded on $\B \times [0, \tau]$, $\szero(\bbeta, t)$ is bounded away from zero.
\item The penalty function satisfies that $a_K=O_p(K^{-1/2})$ and $b_K \to 0$.
\item $p_{\lambda_n}(|\beta|)$ satisfies  $\lim\limits_{K\to \infty}K^{1/2} \inf\limits_{|\beta|\leq CK^{-1/2}}p'_{\lambda_K}(|\beta|) \to \infty$.
\end{enumerate}

The following theorem states selection consistency and the oracle properties of the penalization methods for the marginal PSH model.
\begin{thm}
\label{thm3}
Under Condition B.1 - B.4 in the Appendix, the following holds as $K \to \infty$,
\begin{enumerate}
\item [a.] $\tbeta^M$ is a root-$K$ consistent estimator for $\bbeta_0$, i.e. $||\tbeta^M - \bbeta_0||=O_p(K^{-1/2})$.
\item [b.] (Oracle properties) With probability tending to 1, the root-$K$ consistent estimator $\tbeta$ satisfies
\begin{enumerate}
\item[i.] (Sparsity) $\tbeta^M_{2}=\bzero$
\item[ii.] (Asymptotic normality) $K^{1/2}(\bOmega^M_{11}+\bP_{11})\{\tbeta^M_{1} -\bbeta_{01}+(\bOmega^M_{11}+\bP_{11})^{-1}\bb_1\} \to N(\bzero, \bSigma^M_{11}(\bbeta_0))$, where $\bOmega^M_{11}$ and $\bSigma^M_{11}$ are the first $s \times s$ submatrix of $\bOmega^M(\bbeta_{0})$ and $\bSigma^M(\bbeta_{0})$, defined in the Appendix.
\end{enumerate}
\end{enumerate}
\end{thm}

%%%%%%%%%%%%%%%%%%%%%%%%%%%%%%%%%%%%%%%%%%

\subsection{Proof of Theorem 3.a. }
It is sufficient to prove that for any positive $\epsilon$, there exists a large constant $C$ such that 
\begin{align}
\label{cluster1}
Pr\{\sup_{\|\bu\|=C} Q^M(\bbeta_0+\alpha_K\bu) < Q^M(\bbeta_0)\}\geq 1-\epsilon
\end{align}
where $\alpha_K=K^{-1/2}+a_K$. 
Denote $D_K(\bu)=\dfrac{1}{K}\{Q^C(\bbeta_0+\alpha_K\bu)-Q^C(\bbeta_0)\}$, and 
\begin{align}
\label{high1}
D_K(\bu) &\leq \dfrac{1}{K}l^M(\bbeta_0+\alpha_K\bu)-l^M(\bbeta_0)\}-\sum_{j=1}^s \{p_{\lambda_K}(|\beta_{j0}+\alpha_Ku_j|)-p_{\lambda_K}(|\beta_{j0}|)\}
\end{align}
By Taylor expansion, the first part
\begin{align}
I_1&=\dfrac{1}{K}(\dfrac{\partial l^M(\bbeta_0)}{\partial \bbeta})^T\alpha_K\bu - \dfrac{1}{2}\bu^T\{-\dfrac{1}{K}\dfrac{\partial^2 l^M(\bbeta_0)}{\partial \bbeta^T\partial \bbeta}+o_p(1)\}\bu\alpha_K^2\\
&=O_p(\alpha^2_KC)+O_p(\alpha_K^2C^2)
\end{align}
From Section \ref{them1a}, term $I_2$ has the order of $C\alpha_K^2$ if $b_K\to 0$. 
By choosing a sufficiently large $C$, $O_p(C^2\alpha_K^2)$ dominates the others. Thus inequality(\ref{cluster1}) holds.

%%%%%%%%%%%%%%%%%%%%%%%%%%%%%%%%%%%%%%%%%%%%

\subsection{Proof of Theorem 3.b.(i) }
Define the following process
\begin{align}
X(\bbeta, \tau)&=\dfrac{1}{K}\{l^M(\bbeta)-l^M(\bbeta_0)\}\\
&=\dfrac{1}{K}\sum_{k=1}^K\sum_{i=1}^{n_k}\int_0^{\tau}\{(\bbeta-\bbeta_0)^T\bZ_{ki}-\log\dfrac{\hSzero(\bbeta, u)}{\hSzero(\bbeta_0, u)}\}\hw_{ki}(u)dN_{ki}(u)\\
A(\bbeta, \tau)&=\dfrac{1}{K}\sum_{k=1}^K\sum_{i=1}^{n_k}\int_0^{\tau}\{(\bbeta-\bbeta_0)^T\bZ_{ki}-\log\dfrac{\hSzero(\bbeta, u)}{\hSzero(\bbeta_0, u)}\}\hw_{ki}(u)\lambda_{1}(u)du\\
&=\int_{0}^{\tau}\{(\bbeta-\bbeta_0)^T\hSone(\beta_0, u)-\hSzero(\bbeta_0, u)\log\dfrac{\hSzero(\bbeta, u)}{\hSzero(\bbeta_0, u)}\}\lambda_{10}(u)du
\end{align}
From Zhou et al (2012), we have $X(\bbeta, .)-A(\bbeta, .) \to 0$ in probability and $\lim_{K \to \infty}\hSp(\bbeta, t)=G(t)\spp(\bbeta, t)$. Thus for each $\beta \in \B$, 
\begin{align}
A(\beta, \tau) \to f(\bbeta, \tau)=\int_{0}^{\tau}\{(\bbeta-\bbeta_0)^T\sone(\beta_0, u)-\szero(\bbeta_0, u)\log\dfrac{\szero(\bbeta, u)}{\szero(\bbeta_0, u)}\}\lambda_{10}(u)du
\end{align}
where $f$ has the following properties
\begin{align}
f(\bbeta_0)=0, \ \ \dfrac{\partial f(\bbeta_0)}{\partial \bbeta}=0, \ \ -\dfrac{\partial^2 f(\bbeta_0)}{\partial \bbeta\partial \bbeta^T}=\bOmega^M(\bbeta_0)
\end{align}
Then the following holds 
\begin{align}
\dfrac{1}{K}\{l^M(\bbeta)-l^M(\bbeta_0)\}&=f(\bbeta, \tau) +O_p(\dfrac{||\bbeta-\bbeta_0||}{\sqrt{K}})
\end{align}
We also have $f(\bbeta)=-\dfrac{1}{2}(\bbeta-\bbeta_0)^T\{\Omega(\bbeta_0)+o_p(1)\}(\bbeta-\bbeta_0)$.
By Taylor expansion, we obtain
\begin{align}
\dfrac{\partial Q^M(\bbeta)}{\partial\beta_j}&=\dfrac{\partial l^M(\bbeta)}{\partial \beta_j}-Kp'_{\lambda_K}(|\beta_j|)sgn(\beta_j)\\
&=K\sum^d_{l=1}\dfrac{\partial^2 f(\bbeta_0)}{\partial\beta_j \partial \beta_l}(\beta_l - \beta_{l0}) +O_p(K^{1/2}||\bbeta-\bbeta_0||)-Kp'_{\lambda_K}(|\beta_j|)sgn(\beta_j)\\
&=O_p(K^{1/2})+O_p(1)-Kp'_{\lambda_n}(|\beta_j|)sgn(\beta_j)\\
&=K^{1/2}\{O_p(1)-K^{1/2}p'_{\lambda_n}(|\beta_j|)sgn(\beta_j)\}
\end{align}
Since $\lim\limits_{K\to \infty} K^{1/2} \inf\limits_{||\beta||\leq CK^{-1/2}}p'_{\lambda_K}(|\beta|) \to \infty, k=1, \dots, K$, the sign of $\dfrac{\partial Q^C(\bbeta)}{\beta_j}$ is completely determined by the sign of $\beta_j$. 
%%%%%%%%%%%%%%%%%%%%%%%%%%%%%%%%%%%%%%%%
\subsection{Proof of Theorem 3.b. (ii)}
\label{them1b2}
It is easy to show that there exists a root-n consistent local maximizer $\tbeta_{1}$ of $Q(\bbeta_{1}, \bzero)$, satisfying equation\\
$\dfrac{\partial Q^M(\bbeta)}{\partial\bbeta_{1}}\Big|_{\bbeta=(\tbeta_{1}^T, \bzero)^T}=0$.
Denote $U^M(\bbeta)$ and $H^M(\bbeta)$ as the score function and the Hessian matrix of $l^M(\bbeta)$ and it is
\begin{equation}
U^M(\bbeta)=\sum_{k=1}^K\sum_{i=1}^{n_k}\int_0^{\infty}\{\bZ_{ik}-\dfrac{\hSone(\bbeta, u)}{\hSzero(\bbeta, u)}\}\hw_{ik}(u)dN_{ik}(u)
\end{equation}
\begin{equation}
H^M(\bbeta)=\sum_{k=1}^K\sum_{i=1}^{n_k}\int_0^{\infty}\left [\dfrac{\hStwo(\bbeta, u)}{\hSzero(\bbeta, u)}-\{\dfrac{\hSone(\bbeta, u)}{\hSzero(\bbeta, u)}\}^{\otimes2}\right ]\hw_{ik}(u)dN_{ik}(u)
\end{equation}
For $j=1, \dots, s$
\begin{equation}
\begin{split}
0&=\dfrac{\partial l^M(\bbeta)}{\partial\beta_j}\Big|_{\bbeta=(\tbeta_{1}^T, \bzero)^T}-Kp'_{\lambda_n}(|\tilde\beta_j|)sgn(\tilde\beta_j)\\
&=\dfrac{\partial l^M(\bbeta_0)}{\partial \beta_j}-\sum_{l=1}^s\{-\dfrac{\partial^2 f(\bbeta_0)}{\partial\beta_j \partial \beta_l}+o_p(1)\}(\tilde\beta_l-\beta_{l0})-\\
&\quad  K\left(p'_{\lambda_K}(|\beta_{j0}|)sgn(\beta_{j0})+\{p''_{\lambda_K}(| \beta_{j0}|)+o_p(1)\}(\tilde\beta_j-\beta_{j0})\right)
\end{split}
\end{equation}
Denote $U^M_{11}(\bbeta)$ as the first $s$ elements of the score function $U^M(\bbeta)$. As $K \to \infty$, $K^{-1/2}U^M_{11}(\bbeta_0) \to N(\bzero, \bSigma^M_{11}(\bbeta_0))$ in distribution, where $\bSigma^M_{11}$ is the first $s\times s$ submatrix of $\bSigma^M$. Furthermore, let $H^M_{11}$ be the first $s\times s$ submatrix of of $H^M$ and $-\dfrac{1}{K}H^M_{11} \to \bOmega^M_{11}(\bbeta_0)$ in probability, where $\bOmega^M_{11}$ is the first $s\times s$ submatrix of $\bOmega^M$. Let $\bb_1$ be the first $s$ elements of $\bb$, and $\bP_{11}$ be the first $s \times s$ submatrix of $\bP$. By Slutsky's Theorem, 
\begin{align}
K^{1/2}(\bOmega^M_{11}+\bP_{11})\{\tbeta_{1} -\bbeta_{01}+(\bOmega^M_{11}+\bP_{11})^{-1}\bb_1\} \to N(\bzero, \bSigma^M_{11}(\bbeta_0))
\end{align}
This completes the proof.

To achieve the oracle properties, penalty functions need satisfy Condition $B.3 - B.4$. Borrowing the arguments in Section (\ref{oracle}), we can easily show that the oracle properties of ALASSO, SCAD, and MCP hold when $\lambda$ is chosen appropriately: (a). ALASSO possesses the oracle properties when $\sqrt{K}\lambda_{K} \to 0$, $K\lambda_K \to \infty$, and $\theta_j=|\hbeta^M_j|^{-1}$, where $|\hbeta^M_j|^{-1}$ is the maximizer of the pseudopartial likelihood with root-$K$ consistency; (b). SCAD and MCP have the oracle properties when $\lambda_K \to 0$ and $\sqrt{K}\lambda_K \to \infty$; (c). the oracle properties do not hold for  LASSO.

\end{appendices}

\begin{appendices}
\section{Complete Selection Results in the UNOS Data Analysis}
\begin{landscape}
\begin{table*}
\caption{Complete Variable selection results and estimated regression coefficients using UNOS data. }
\scalebox{0.45}{
\begin{threeparttable}
\begin{tabular}{llllllllllllll}
\hline
  &  & \multicolumn{4}{c}{Regularly Stratified} & \multicolumn{4}{c}{Highly Stratified} & \multicolumn{4}{c}{Marginal} \\
& KDRI  & LASSO & ALASSO & SCAD & MCP & LASSO & ALASSO & SCAD & MCP & LASSO & ALASSO & SCAD & MCP \\
\hline
{\bf Donor Factors} & &  &  &  &  &  &  &  &  &  &  &  &  \\
Age -40 yrs & 0.013(0.002) & 0.012(0.001) & 0.013(0.001) & 0.013(0.001) & 0.012(0.001) & 0.011(0.001) & 0.013(0.001) & 0.013(0.001) & 0.012(0.001) & 0.012(0.001) & 0.013(0.001) & 0.012(0.001) & 0.012(0.001) \\
(applies to all ages)  & &  &  &  &  &  &  &  &  &  &  &  &  \\
Age -18 yrs & -0.019(0.010) & -0.01(0.008) & -0.013(0.006) & -0.005(0.007) & -0.005(0.007) & -0.015(0.006) & -0.014(0.005) & -0.005(0.005) & -0.005(0.005) & -0.014(0.008) & -0.018(0.007) & -0.014(0.008) & -0.014(0.008) \\
(applies only if age\textless18)  & &  &  &  &  &  &  &  &  &  &  &  &  \\
Age -50 yrs & 0.011(0.006) & 0.020(0.003) & 0.018(0.003) & 0.018(0.003) & 0.019(0.003) & 0.018(0.003) & 0.016(0.002) & 0.018(0.002) & 0.019(0.002) & 0.020(0.003) & 0.018(0.003) & 0.020(0.003) & 0.020(0.003) \\
(applies only if age\textgreater50)  & &  &  &  &  &  &  &  &  &  &  &  &  \\
Afrian American race & 0.179(0.071) & 0.170(0.031) & 0.167(0.031) & 0.172(0.031) & 0.168(0.031) & 0.151(0.023) & 0.148(0.027) & 0.174(0.026) & 0.169(0.027) & 0.164(0.033) & 0.157(0.033) & 0.166(0.033) & 0.166(0.033) \\
Male & - & -0.037(0.026) & - & - & - & -0.042(0.018) & - & - &-  & -0.035(0.026) & -0.027(0.023) & -0.034(0.026) & -0.034(0.026) \\
Height: per 10 cm increase & -0.046(0.015) & -0.021(0.013) &  & -0.031(0.011) & -0.029(0.011) & -0.008(0.01) &  & -0.031(0.008) & -0.028(0.009) & -0.018(0.014) & - & -0.019(0.014) & -0.019(0.014) \\
Weight & -0.020(0.010) & -0.227(0.06) & -0.311(0.052) & -0.248(0.059) & -0.241(0.059) & -0.169(0.048) & -0.215(0.043) & -0.251(0.05) & -0.244(0.045) & -0.213(0.064) & -0.235(0.059) & -0.220(0.064) & -0.220(0.064) \\
(per 5 kg increase if \textless 80 kg)  & &  &  &  &  &  &  &  &  &  &  &  &  \\
Donation after cardiac death & 0.133(0.132) & - & - & - & - & - & - & - & - & - & - & - & - \\
Cause of  death: stroke & 0.088(0.051) & 0.061(0.025) & 0.073(0.024) & - & 0.067(0.024) & 0.067(0.021) & 0.040(0.019) & - & 0.069(0.022) & 0.066(0.025) & 0.053(0.025) & 0.066(0.025) & 0.066(0.025) \\
Serum creatinine-1& 0.220(0.051) & 0.195(0.036) & 0.174(0.035) & 0.183(0.035) & 0.186(0.035) & 0.150(0.026) & 0.107(0.028) & 0.183(0.022) & 0.187(0.029) & 0.192(0.035) & 0.155(0.035) & 0.196(0.035) & 0.196(0.035) \\
(applies to all Cr values )  & &  &  &  &  &  &  &  &  &  &  &  &  \\
Serum creatinine-1 & -0.209(0.082) & -0.189(0.05) & -0.165(0.049) & -0.176(0.049) & -0.179(0.049) & -0.143(0.034) & -0.099(0.036) & -0.177(0.032) & -0.18(0.036) & -0.185(0.046) & -0.148(0.046) & -0.189(0.046) & -0.189(0.046) \\
(applies if Cr\textgreater1.5 )  & &  &  &  &  &  &  &  &  &  &  &  &  \\
Diabetic & 0.130(0.076) & 0.233(0.047) & 0.237(0.047) & 0.228(0.047) & 0.232(0.047) & 0.212(0.039) & 0.191(0.038) & 0.231(0.034) & 0.235(0.036) & 0.242(0.05) & 0.229(0.049) & 0.245(0.05) & 0.245(0.05) \\
Hypertensive & 0.126(0.102) & 0.137(0.026) & 0.139(0.026) & 0.151(0.026) & 0.138(0.026) & 0.132(0.02) & 0.130(0.021) & 0.15(0.018) & 0.136(0.022) & 0.142(0.027) & 0.143(0.027) & 0.142(0.027) & 0.142(0.027) \\
Cigarette users & - & 0.027(0.021) & - & - & - & 0.017(0.013) & - & - & - & 0.024(0.018) & - & 0.025(0.018) & 0.025(0.018) \\
Positive HCV status & 0.240(0.153) & 0.098(0.068) & - & - & - & 0.070(0.048) & - & - & - & 0.133(0.064) & 0.055(0.064) & 0.135(0.064) & 0.135(0.064) \\
{\bf Transplant Factors} &  &  &  &  &  &  &  &  &  &  &  &  &  \\
Cold ischemic time& 0.005(0.003) & 0.003(0.001) & - & - & - & 0.002(0.001) & -& - & - & 0.004(0.002) & 0.002(0.001) & 0.004(0.002) & 0.004(0.002) \\
( per 1 hr (ref=20 hr) )  & &  &  &  &  &  &  &  &  &  &  &  &  \\
Organ sharing &  &  &  &  &  &  &  &  &  &  &  &  &  \\
Regional(ref=local) & - & 0.061(0.037) & - & - & - & 0.045(0.022) & - & - & - & 0.030(0.038) & - & 0.031(0.038) & 0.031(0.038) \\
National(ref=local) & - & -0.052(0.031) & - & - & - & -0.038(0.027) & - & - & - & -0.049(0.037) & - & -0.051(0.037) & -0.051(0.037) \\
HLA-B mismatch &  &  &  &  &  &  &  &  &  &  &  &  &  \\
\ \ 0 (ref=2 B MM) & -0.077(0.056) & -0.155(0.034) & -0.176(0.031) & -0.174(0.031) & -0.174(0.031) & -0.141(0.028) & -0.134(0.025) & -0.174(0.027) & -0.174(0.03) & -0.154(0.039) & -0.158(0.039) & -0.156(0.039) & -0.156(0.039) \\
\ \ 1 & -0.061(0.041) & -0.068(0.023) & -0.07(0.023) & -0.069(0.023) & -0.069(0.023) & -0.057(0.019) & -0.051(0.021) & -0.067(0.018) & -0.067(0.017) & -0.055(0.025) & -0.052(0.024) & -0.056(0.025) & -0.056(0.025) \\
HLA-DR mismatch &  &  &  &  &  &  &  &  &  &  &  &  &  \\
\ \ 0 (ref=1 DR MM) & -0.130(0.041) & -0.097(0.027) & -0.103(0.027) & -0.102(0.027) & -0.102(0.027) & -0.087(0.024) & -0.078(0.02) & -0.102(0.024) & -0.102(0.023) & -0.091(0.03) & -0.084(0.029) & -0.092(0.03) & -0.092(0.03) \\
\ \ 2 & 0.077(0.051) & 0.055(0.023) & 0.057(0.023) & 0.058(0.023) & 0.057(0.023) & 0.051(0.018) & 0.047(0.019) & 0.058(0.017) & 0.057(0.02) & 0.063(0.025) & 0.058(0.025) & 0.063(0.025) & 0.063(0.025) \\
Transplant year & - & -0.155(0.021) & -0.161(0.021) & -0.165(0.021) & -0.163(0.021) & -0.139(0.018) & -0.139(0.021) & -0.166(0.02) & -0.164(0.02) & -0.151(0.025) & -0.156(0.025) & -0.152(0.025) & -0.151(0.025) \\
En bloc transplant & -0.346(0.138) & -0.313(0.117) & -0.285(0.116) & -0.317(0.117) & -0.312(0.117) & -0.209(0.094) & -0.083(0.085) & -0.310(0.094) & -0.304(0.081) & -0.295(0.124) & -0.132(0.122) & -0.304(0.124) & -0.304(0.124) \\
Double kidney transplant & -0.148(0.22) & -0.334(0.093) & -0.329(0.092) & -0.332(0.092) & -0.33(0.092) & -0.247(0.063) & -0.181(0.047) & -0.331(0.067) & -0.330(0.061) & -0.329(0.078) & -0.250(0.078) & -0.338(0.078) & -0.338(0.078) \\
ABO identical & - & - & - & - & - & - & - & - & - & -0.023(0.052) & - & -0.026(0.052) & -0.027(0.052) \\
{\bf Recipient Factors} &  &  &  &  &  &  &  &  &  &  &  &  &  \\
Age &  & -0.025(0.001) & -0.025(0.001) & -0.025(0.001) & -0.025(0.001) & -0.024(0.001) & -0.025(0.001) & -0.025(0.001) & -0.025(0.001) & -0.025(0.001) & -0.025(0.001) & -0.025(0.001) & -0.025(0.001) \\
African America &  & 0.356(0.025) & 0.359(0.025) & 0.363(0.024) & 0.363(0.024) & 0.362(0.02) & 0.386(0.02) & 0.364(0.02) & 0.364(0.019) & 0.372(0.027) & 0.384(0.028) & 0.371(0.027) & 0.371(0.027) \\
Male &  & -0.049(0.022) & -0.053(0.022) & - & - & -0.032(0.017) & -0.018(0.015) & - & - & -0.06(0.026) & -0.039(0.021) & -0.063(0.026) & -0.063(0.026) \\
Primary diagnosis (ref=GN) &  &  &  &  &  &  &  &  &  &  &  &  &  \\
\ \ Diabetes &  & -0.075(0.03) & -0.080(0.03) & -0.079(0.03) & -0.079(0.03) & -0.063(0.023) & -0.053(0.023) & -0.074(0.023) & -0.074(0.026) & -0.059(0.033) & -0.054(0.033) & -0.060(0.033) & -0.060(0.033) \\
\ \ Hypertension &  & 0.102(0.029) & 0.100(0.029) & 0.098(0.029) & 0.097(0.029) & 0.088(0.02) & 0.074(0.021) & 0.101(0.019) & 0.101(0.021) & 0.119(0.028) & 0.109(0.028) & 0.121(0.028) & 0.121(0.028) \\
\ \ Failed transplants &  & 0.545(0.142) & 0.546(0.142) & 0.538(0.142) & 0.539(0.142) & 0.479(0.116) & 0.432(0.096) & 0.540(0.094) & 0.542(0.109) & 0.561(0.141) & 0.522(0.141) & 0.570(0.141) & 0.570(0.141) \\
\ \ CAKUT/Congenital uropathy &  & -0.191(0.038) & -0.193(0.038) & -0.191(0.038) & -0.192(0.038) & -0.164(0.03) & -0.136(0.03) & -0.194(0.029) & -0.194(0.029) & -0.197(0.04) & -0.179(0.039) & -0.200(0.04) & -0.200(0.04) \\
\ \ Others &  & 0.048(0.032) & 0.051(0.032) & 0.057(0.032) & 0.057(0.032) & 0.045(0.032) & 0.044(0.025) & 0.061(0.023) & 0.061(0.027) & 0.072(0.037) & 0.072(0.037) & 0.073(0.037) & 0.073(0.037) \\
Blood transfusion &  & 0.040(0.022) & - & - & - & 0.034(0.017) & - & - & - & 0.032(0.023) & - & 0.033(0.023) & 0.033(0.023) \\
Height &  & - & - & - & - & - & - & - & - & 0.001(0.001) & - & 0.001(0.001) & 0.001(0.001) \\
Weight &  & 0.005(0.001) & 0.005(0.001) & 0.005(0.001) & 0.005(0.001) & 0.004(0.0005) & 0.004(0.001) & 0.005(0.0005) & 0.005(0.0005) & 0.005(0.001) & 0.005(0.001) & 0.005(0.001) & 0.005(0.001) \\
Peak PRA (ref=0) &  &  &  &  &  &  &  &  &  &  &  &  &  \\
\ \ 1-50 &  & 0.057(0.023) & 0.057(0.023) & 0.060(0.023) & 0.061(0.023) & 0.044(0.015) & 0.029(0.019) & 0.060(0.02) & 0.061(0.017) & 0.037(0.024) & 0.031(0.024) & 0.038(0.024) & 0.038(0.024) \\
\ \ 51-80 &  & 0.116(0.057) & 0.119(0.057) & 0.138(0.057) & 0.137(0.057) & 0.095(0.048) & 0.066(0.047) & 0.142(0.047) & 0.141(0.048) & 0.096(0.06) & 0.081(0.06) & 0.098(0.06) & 0.098(0.06) \\
\ \ \textgreater80 &  & 0.244(0.048) & 0.246(0.047) & 0.268(0.047) & 0.267(0.047) & 0.196(0.035) & 0.134(0.036) & 0.269(0.037) & 0.268(0.034) & 0.231(0.049) & 0.194(0.049) & 0.236(0.049) & 0.236(0.049) \\
Years of RRT (ref: \textless=1) &  &  &  &  &  &  &  &  &  &  &  &  &  \\
\ \ 2-3 &  & - & - & - & - & - & - & - & - & -0.013(0.022) & -0.004(0.022) & -0.015(0.022) & -0.015(0.022) \\
\ \ \textgreater3 &  & - & - & - & - & - & - & - & - & -0.031(0.028) & -0.010(0.027) & -0.035(0.028) & -0.035(0.028) \\
Angina pectoris &  & -0.059(0.035) & - & - & - & -0.041(0.026) &  &  &  & -0.048(0.035) & -0.034(0.035) & -0.050(0.035) & -0.050(0.035) \\
Peripheral vascular disease &  & - & - & - & - & - & - & - & - & - & - & - & - \\
COPD &  & - & - & - & - & - & - & - & - & -0.085(0.115) & - & -0.092(0.115) & -0.092(0.115) \\
Positive HCV status &  & 0.207(0.044) & 0.242(0.037) & 0.237(0.037) & 0.238(0.037) & 0.196(0.035) & 0.208(0.032) & 0.237(0.026) & 0.237(0.029) & 0.195(0.043) & 0.204(0.043) & 0.197(0.043) & 0.197(0.043) \\
Diabetic &  & - & - & - & - & - & - & - & - & - & - & - & -\\
\hline
\end{tabular}
\begin{tablenotes}
\item HCV: Hepatitis C.
Cr: serum creatinine.
 HLA: human leukocyte antigen.
GN: glomerulonephritis
CAKUT: congenital anomalies of the kidney and urinary tract.
RRT: renal replacement therapy
COPD: chronic obstructive pulmonary disease.
  \end{tablenotes}
 \end{threeparttable}
}
\end{table*}
\end{landscape}
\end{appendices}

\begin{appendices}
\section{Simulations Results for Dependent Censoring}

We consider three scenarios when censoring times and covariates are dependent. In Scenario (a), we allowed the censoring to depend on important variables $z_1$ and $z_4$; in Scenario (b), censoring times depend on non-important variables $z_5$ and $z_8$; in Scenario (c), censoring times depend on both important and non-important variables $z_1$ and $z_3$; in Scenario (d), we displayed the results when censoring and covariates are independent.
\subsection{Penalized Stratified PSH Model}
\subsubsection{Regularly Stratified}
\begin{table}[!h]
\centering
\caption{Selection results of the penalized stratified based on 100 replications with different censoring patterns, where $K=3$, $n=200$, $p=0.6$, the censoring rate is 28\%, and the event rate is 45\%}
\scalebox{0.9}{
\begin{tabular}{lllllll}
\hline\noalign{\smallskip}
Scenario & Depend on   & Penalty & C (5) & IC (0) & Pcorr & MMSE  \\
\noalign{\smallskip}\hline\noalign{\smallskip}
(a)        & $z_1$, $z_4$      & MPLE    & 0     & 0      & 0\%   & 0.137 \\
         &             & LASSO   & 3.43  & 0      & 20\%  & 0.166 \\
         &             & ALASSO  & 4.73  & 0.02   & 76\%  & 0.086 \\
         &             & SCAD    & 4.89  & 0.03   & 90\%  & 0.053 \\
         &             & MCP     & 4.90   & 0.02   & 90\%  & 0.052 \\
         &             & Oracle  & 5     & 0      & 100\% & 0.050 \\
         \noalign{\smallskip}\hline\noalign{\smallskip}
(b)        & $z_5$, $z_8$      & MPLE    & 0     & 0      & 0\%   & 0.143 \\
         &             & LASSO   & 3.64  & 0.01   & 21\%  & 0.125 \\
         &             & ALASSO  & 4.79  & 0.03   & 78\%  & 0.084 \\
         &             & SCAD    & 4.89  & 0.02   & 87\%  & 0.059 \\
         &             & MCP     & 4.92  & 0.02   & 90\%  & 0.056 \\
         &             & Oracle  & 5     & 0      & 100\% & 0.051 \\
         \noalign{\smallskip}\hline\noalign{\smallskip}
(c)        &$z_1$, $z_3$      & MPLE    & 0     & 0      & 0\%   & 0.134 \\
         &             & LASSO   & 3.54  & 0      & 23\%  & 0.148 \\
         &             & ALASSO  & 4.79  & 0.02   & 81\%  & 0.083 \\
         &             & SCAD    & 4.88  & 0.02   & 89\%  & 0.056 \\
         &             & MCP     & 4.90   & 0.02   & 91\%  & 0.054 \\
         &             & Oracle  & 5     & 0      & 100\% & 0.050 \\
                      \noalign{\smallskip}\hline\noalign{\smallskip}
(d)        & Independent & MPLE    & 0     & 0      & 0\%   & 0.133 \\
         &             & LASSO   & 3.37  & 0      & 23\%  & 0.112 \\
         &             & ALASSO  & 4.75  & 0.01   & 79\%  & 0.064 \\
         &             & SCAD    & 4.92  & 0      & 92\%  & 0.052 \\
         &             & MCP     & 4.92  & 0      & 92\%  & 0.056 \\
         &             & Oracle  & 5     & 0      & 100\% & 0.044\\
         \noalign{\smallskip}\hline
\end{tabular}}
\end{table}

\subsubsection{Highly Stratified}

\begin{table}[h]
\centering
\caption{Selection results of the penalized stratified model based on 100 replications  with different censoring patterns, where $\alpha_1=0.7$, $K=100$, $n_k \in \{2, 3, 4, 5\}$. The censoring rate is 27\% and the event rate is 46\%.}
\begin{tabular}{lllllll}
\hline\noalign{\smallskip}
Scenario & Depend on   & Penalty & C (5) & IC (0) & Pcorr & MMSE  \\
\noalign{\smallskip}\hline\noalign{\smallskip}
(a)        & $z_1$, $z_4$     & MPLE    & 0    & 0    & 0\%   & 0.168 \\
         &             & LASSO   & 3.60  & 0.01 & 20\%  & 0.200 \\
         &             & ALASSO  & 4.82 & 0.04 & 81\%  & 0.113 \\
         &             & SCAD    & 4.92 & 0.09 & 85\%  & 0.070 \\
         &             & MCP     & 4.91 & 0.09 & 86\%  & 0.070 \\
         &             & Oracle  & 5    & 0    & 100\% & 0.066 \\
                               \noalign{\smallskip}\hline\noalign{\smallskip}
(b)        &$z_5$, $z_8$      & MPLE    & 0    & 0    & 0\%   & 0.167 \\
         &             & LASSO   & 3.73 & 0.01 & 26\%  & 0.199 \\
         &             & ALASSO  & 4.76 & 0.04 & 77\%  & 0.124 \\
         &             & SCAD    & 4.88 & 0.11 & 83\%  & 0.070 \\
         &             & MCP     & 4.88 & 0.13 & 83\%  & 0.070 \\
         &             & Oracle  & 5    & 0    & 100\% & 0.063 \\
                               \noalign{\smallskip}\hline\noalign{\smallskip}
(c)        & $z_1$, $z_3$  & MPLE    & 0    & 0    & 0\%   & 0.174 \\
         &             & LASSO   & 3.59 & 0.01 & 29\%  & 0.185 \\
         &             & ALASSO  & 4.73 & 0.06 & 75\%  & 0.112 \\
         &             & SCAD    & 4.92 & 0.10  & 86\%  & 0.069 \\
         &             & MCP     & 4.93 & 0.10  & 85\%  & 0.069 \\
         &             & Oracle  & 5    & 0    & 100\% & 0.065 \\
                               \noalign{\smallskip}\hline\noalign{\smallskip}
(d)        & Independent & MPLE    & 0    & 0    & 0\%   & 0.164 \\
         &             & LASSO   & 3.61 & 0    & 22\%  & 0.178 \\
         &             & ALASSO  & 4.76 & 0.02 & 79\%  & 0.097 \\
         &             & SCAD    & 4.90  & 0.06 & 88\%  & 0.060 \\
         &             & MCP     & 4.90  & 0.05 & 87\%  & 0.060 \\
         &             & Oracle  & 5    & 0    & 100\% & 0.058\\
                  \noalign{\smallskip}\hline
\end{tabular}
\end{table}

\newpage
\subsection{Penalized Marginal PSH Model}
\begin{table}[h]
\centering
\caption{Selection results of the penalized marginal model based on 100 replications with different censoring patterns, where $\alpha_1=0.7$, $K=100$, $n_k \in \{2, 3, 4, 5\}$. The censoring rate is 29\% and the event rate is 43\%.}
\begin{tabular}{lllllll}
\hline\noalign{\smallskip}
Scenario & Depend on   & Penalty & C (5) & IC (0) & Pcorr & MMSE  \\
\noalign{\smallskip}\hline\noalign{\smallskip}
(a)        &  $z_1$, $z_4$      & MPLE    & 0     & 0      & 0\%   & 0.101 \\
         &             & LASSO   & 3.36  & 0      & 15\%  & 0.065 \\
         &             & ALASSO  & 4.77  & 0      & 83\%  & 0.039 \\
         &             & SCAD    & 4.82  & 0      & 86\%  & 0.034 \\
         &             & MCP     & 4.81  & 0      & 86\%  & 0.033 \\
         &             & Oracle  & 5     & 0      & 100\% & 0.031 \\
                  \noalign{\smallskip}\hline\noalign{\smallskip}
(b)        &  $z_5$, $z_8$      & MPLE    & 0     & 0      & 0\%   & 0.108 \\
         &             & LASSO   & 3.64  & 0      & 29\%  & 0.067 \\
         &             & ALASSO  & 4.75  & 0      & 82\%  & 0.033 \\
         &             & SCAD    & 4.80   & 0      & 86\%  & 0.032 \\
         &             & MCP     & 4.83  & 0      & 87\%  & 0.033 \\
         &             & Oracle  & 5     & 0      & 100\% & 0.031 \\
                  \noalign{\smallskip}\hline\noalign{\smallskip}
(c)        &  $z_1$, $z_3$      & MPLE    & 0     & 0      & 0\%   & 0.104 \\
         &             & LASSO   & 3.58  & 0      & 22\%  & 0.071 \\
         &             & ALASSO  & 4.80   & 0      & 84\%  & 0.038 \\
         &             & SCAD    & 4.83  & 0      & 88\%  & 0.034 \\
         &             & MCP     & 4.84  & 0      & 88\%  & 0.034 \\
         &             & Oracle  & 5     & 0      & 100\% & 0.032 \\
                  \noalign{\smallskip}\hline\noalign{\smallskip}
(d)        & Independent & MPLE    & 0     & 0      & 0\%   & 0.097 \\
         &             & LASSO   & 3.56  & 0      & 22\%  & 0.060 \\
         &             & ALASSO  & 4.84  & 0      & 85\%  & 0.035 \\
         &             & SCAD    & 4.88  & 0      & 91\%  & 0.031 \\
         &             & MCP     & 4.84  & 0      & 85\%  & 0.032 \\
         &             & Oracle  & 5     & 0      & 100\% & 0.029\\
                  \noalign{\smallskip}\hline
\end{tabular}
\end{table}

\end{appendices}

\end{document}